\documentclass[a4paper,fleqn]{cas-dc}

\usepackage[numbers]{natbib}
\usepackage{xcolor}
\definecolor{BrickRed}{RGB}{132,31,39}
\definecolor{headcolor}{RGB}{0,0,0}

\def\tsc#1{\csdef{#1}{\textsc{\lowercase{#1}}\xspace}}
\tsc{WGM}
\tsc{QE}
\begin{document}
\let\WriteBookmarks\relax
\def\floatpagepagefraction{1}
\def\textpagefraction{.001}

\shorttitle{}    

\shortauthors{T. G. Seidel et al.}  

%\title [mode = title]{Phase space dynamics of multi-pulse configurations in semiconductor ring lasers}  
\title [mode = title]{Coherent pulse interactions in mode-locked semiconductor lasers}  




\author[1]{Thomas G. Seidel}%
%\email{Author@institution.edu.}

\author[3]{Julien Javaloyes}

\author[1,2]{Svetlana V. Gurevich}
\cormark[1]

\ead{gurevics@uni-muenster.de}

\affiliation[1]{organization={Institute for Theoretical Physics, University of Münster},
	            addressline={Wilhelm-Klemm-Str. 9}, 
	            city={Münster},
	%          citysep={}, % Uncomment if no comma needed between city and postcode
	            postcode={48149}, 
	            country={Germany}}
            
\affiliation[2]{organization={Center for Nonlinear Science (CeNoS), University of Münster},
	addressline={Corrensstrasse 2}, 
	city={Münster},
	%          citysep={}, % Uncomment if no comma needed between city and postcode
	postcode={48149}, 
	country={Germany}}  

\affiliation[3]{organization={Departament de Física, Universitat de les Illes Balears, \& Institute of Applied Computing and Community Code (IAC-3)},
	addressline={C/ Valldemossa km 7.5}, 
	city={Mallorca},
	%          citysep={}, % Uncomment if no comma needed between city and postcode
	postcode={07122}, 
	country={Spain}}

\cortext[1]{Corresponding author}

\begin{abstract}
We study the dynamics of multipulse solutions in mode-locked lasers in presence of time-delayed feedback stemming, e.g., from reflections upon optical elements, and carrier dynamics. We demonstrate that the dynamics of such a high dimensional problem can be successfully described by some effective equations of motion for the pulses' phases and positions. Analyzing the reduced vector field permits disclosing a highly complex dynamics where coherent and incoherent interactions compete. The latter lead to regimes in which pulses can be equidistant or non-equidistant and also have different phase relations. Multi-stability between regimes is also observed as well as emerging limit cycles and global heteroclinic bifurcations in the reduced phase space.
\end{abstract}


\begin{highlights}
\item Equations of motion for interacting mode-locked pulses
\item Temporal localized states
\item Bifurcation analysis
\item Time-delayed feedback
\end{highlights}

%\nocite{*}

% Keywords

\maketitle

Mode-locking (ML) is a well-established method for achieving ultrashort optical pulses with high repetition rates~\cite{H-JSTQE-00} and optical frequency combs~\cite{CY-RMP-03,D-JOSAB-10,PPR-PR-18} which has proven to be highly relevant for a variety of applications ranging from metrology to medical imaging, see e.g.,~\cite{NAW-MTT-95,K-OL-07,LMW-SCI-17,keller96,lorenser04,UHH-NAT-02,Keller2003,AJ-BOOK-17}. Among other ML techniques, passive mode-locking is achieved by combining two elements, a laser amplifier providing gain and a nonlinear loss element, usually a saturable absorber.
It was shown~\cite{MJB-PRL-14} that when a passive mode-locked laser is operated in the long-cavity regime in which the cavity round-trip is much larger than the gain recovery time, the optical pulses become individually addressable temporally localized structures coexisting with the off solution~\cite{MJC-JSTQE-15,CJM-PRA-16,CSV-OL-18}.
There, a very large number of solutions with different number of pulses per round-trip and different arrangement become multistable. 
Besides the fundamental mode-locked regime where a single pulse circulates in the laser cavity, a mode-locked laser can be operated in the harmonic mode-locked regime where the laser cavity supports a train of several pulses within one round-trip. The gain depletion and subsequent recovery in an harmonic mode-locked laser provides an effective repulsion force leading to the formation of a regime with equally spaced pulses~\cite{KCB-JQE-98,NRV-PD-06,JCM-PRL-16}. However, even in this ordered state, it was recently demonstrated that the many pulses emitted by an harmonically mode-locked laser are not necessarily coherent with each other~\cite{SBG-PRL24}; The derived effective equations of motion (EOM) for the phase differences of the neighbouring pulses indicated that the ensemble behaves like coupled oscillators which have multiple equilibria that can be linked to the splay states of the Kuramoto model with short range interactions.
 
\begin{figure}[b!]
	\centering
	\includegraphics[width=1\columnwidth]{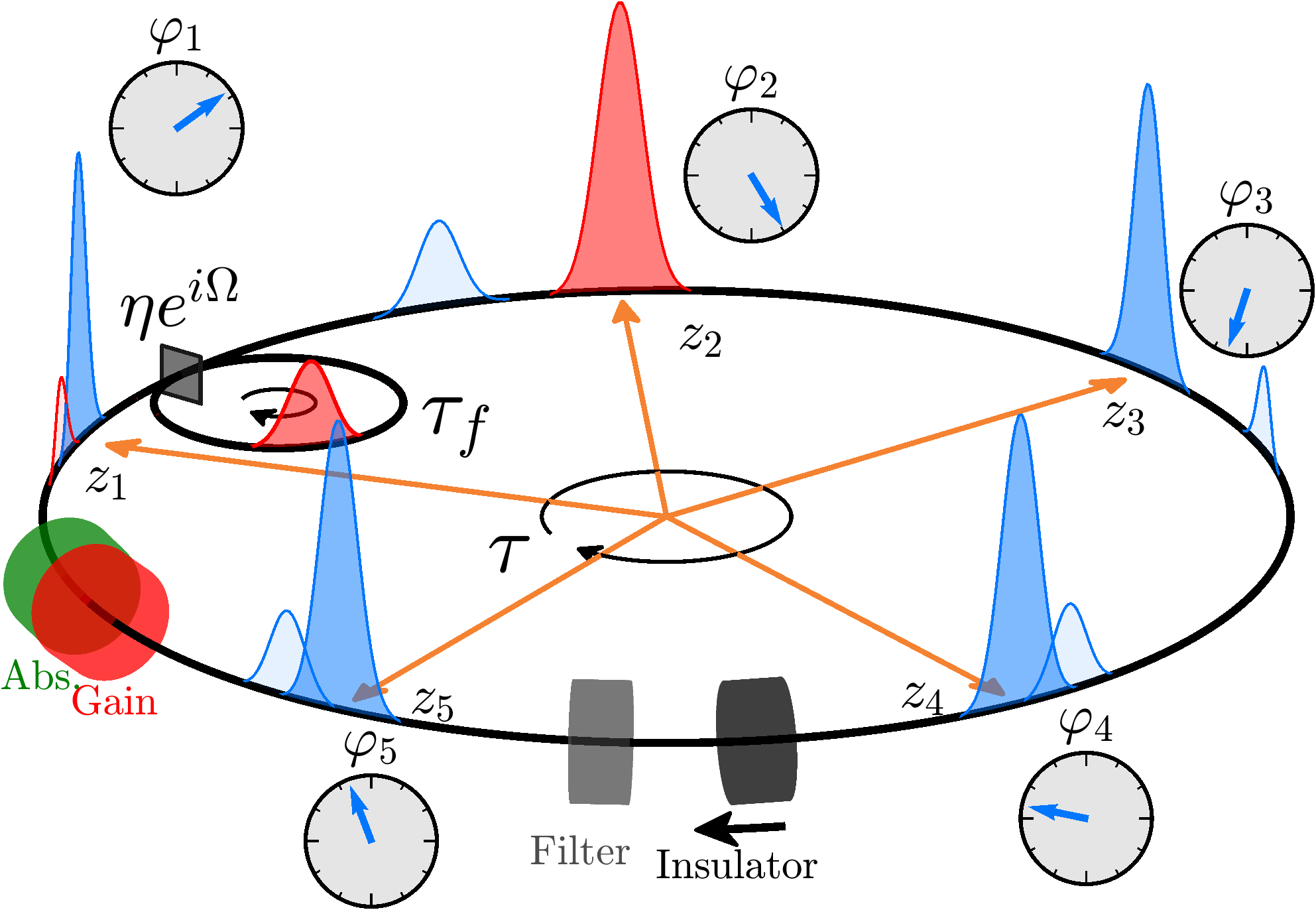}
	\caption{Schematic of a unidirectional ring cavity of length $\tau$. The cavity is mode-locked by a saturable absorber and coupled to a time-delayed feedback loop with round-trip time $\tau_{f}$, feedback rate $\eta$ and phase $\Omega$. Each pulse circulating possesses a phase $\varphi_i$ and position $z_i$. The pulses are non-equidistant and their respective satellites induced by time-delayed feedback exhibit different amount of overlap with neighboring pulses.
	}
	\label{fig:sketch}
\end{figure}

The current manuscript discusses how minute optical feedback stemming from parasitic elements such as intra-cavity lenses or from an external mirror can drastically modify the emission of passively mode-locked lasers in the multi-pulse regimes observed in the long cavity limit.
As a harmonically mode-locked laser is a phase-invariant system, each pulse possesses its own phase and position within the cavity round-trip. As such, a minimal model for the dynamics should be characterized by two degrees of freedom for each pulse: position and phase. Note that these quantities are not necessarily independent as the pulses have several possibilities to interact with each other via carrier dynamics and/or their coherent overlapping due to optical feedback or tail interactions. Since both mechanisms are inherently asymmetric, the resulting interactions are in general non-reciprocal~\cite{FHL-NAT-21}.
First, we focus our analysis on the simplest case of two pulses circulating the cavity in the regime of temporal localized states. Starting with an established Haus master equation model, we derive a set of EOM for the positions and phases of the pulses.
Our effective theory allows us to disclose highly complex dynamics that would be hardly accessible by other means.
In particular, we demonstrate the existence of regimes in which pulses can be equidistant or non-equidistant and also have different phase relations. Multistability between regimes is also observed as well as emerging limit cycles and global heteroclinic bifurcations. From the intuition gained by studying the regimes with two pulses we infer the dynamics observed in the many pulses regimes where clustering, oscillations and chaos can readily be observed. Finally, we compare the results of the EOM with those of the full HME model showing excellent quantitative agreement.

Mode-locking can be experimentally implemented using mirrors to connect the various optical elements and to control the beam shape and the cavity stability. Alternatively, lenses may perform similar functions. However, even with high-quality lenses using anti-reflection coating, minute amounts of optical feedback occur thereby leading to time-delayed feedback perturbations. Notice that time-delayed feedback can also be applied on purpose to control the rich nonlinear passive ML dynamics, e.g., improving the time jitter in high repetition rate mode-locked lasers, controlling the pulse train repetition rate or acting as a solution selector that either reinforces or hinders the appearance of one of the multistable harmonic ML solutions~\cite{OLV-NJP-12,AKB-APL-13,JPR-PRA-15, NJD-OE-16,JKL-CHA-17,BSV-OL-21,SJG-Chaos-22}. The presence of time-delayed feedback in a mode-locked laser creates a non-local coupling between pulses~\cite{JMG-PRL-17} resulting in smaller copies, or satellites, of the main pulses to appear throughout the cavity. Depending on the satellites position they can lead to instabilities of the harmonic mode-locked state~\cite{BSV-OL-21,SJG-Chaos-22} and enable the exchange of phase information between pulses over distances much larger than their typical pulse width. In particular, this can lead to a situation where the phase and position dynamics occurs on the same time scale such that neither of them are enslaved.

A sketch of the system in question is presented in Fig.~\ref{fig:sketch}. Here, we consider a passively mode-locked unidirectional ring laser with round-trip time $\tau$ that contains a gain section and a saturable absorber that enables pulsed emission. Further, a linear bandpass filter ensures that only certain laser modes are amplified, while others are suppressed whereas an optical isolator ensures unidirectional propagation. Additional couplings arising from time-delayed feedback~\cite{BSV-OL-21,SJG-Chaos-22}, or intra-cavity lenses reflection are modeled by an external time-delayed feedback loop of the length $\tau_f$, the feedback rate $\eta$, and the feedback phase $\Omega$. As was mentioned above, time-delayed feedback creates a small satellite for each pulse, which position depends on the ratio $\tau/\tau_f$, e.g. the red colored pulse in Fig.~\ref{fig:sketch} creates the red satellite. Note that the pulses shown in Fig.~\ref{fig:sketch} are not equidistant such that the overlap between pulses and satellites differs.

The interaction of pulses in mode-locked lasers was investigated in detail in the last decades using both experimental, theoretical and numerical tools in e.g., fiber laser cavities~\cite{GA-NAP-12,L-PRA-11,ART-OC-01,WNC-NatCom-19}, solid-state laser with slow saturable absorber~\cite{SCA-JOSAB-99}, figure-of-eight mode-locked fiber laser~\cite{KKK-OLT-20}, optical loop mirror–nonlinear amplifying loop mirror mode-locked laser~\cite{V-PRE-22}, passively mode-locked vertical-cavity surface-emitting lasers~\cite{CJM-PRA-16} or an array of nearest-neighbor coupled passively mode-locked lasers~\cite{PVP-PRL-17}.
Recently, weak interactions of well-separated temporal localized states in semiconductor laser mode-locked by a saturable absorber and operated in the long cavity regime was studied analytically and numerically using a delay differential equations model~\cite{V-Optics-23}. With the aid of the EOM for the position- and phase-differences it was demonstrated that different configurations are obtained by a balance of gain and absorber interactions which depend on the linewidth-enhancement factors.
In particular, it was discussed that the pulse interaction can be separated into incoherent long-range interaction mediated by carriers dynamics and coherent short-range interaction via pulse overlapping tails ~\cite{V-PRE-22,V-Optics-23}. Here, we focus on the influence of the coherent long-range interaction enabled by the presence of time-delayed feedback.
We employ the Haus master equation (HME)~\cite{H-JSTQE-00} to model the optical field in the longitudinal direction while assuming a constant transversal mode. The HME can be  derived from the time-delayed description in \cite{VT-PRA-05} using the assumptions of small gain, losses, and weak spectral filtering~\cite{KNE-PD-06}. This leads to the coupled evolution of the electric field ($E$), the gain ($g$) and absorber population ($q$),

\begin{align}
		\begin{split}
			\partial_{\xi}E= & \left(\frac{1}{2\gamma^{2}}\partial_{z}^{2}+\frac{1-i\alpha_g}{2}g-\frac{1-i\alpha_q}{2}q-k\right)E\\
			&+\eta e^{i\Omega} E\left(z-\tau_{f}\right),\label{eq:Haus1}
		\end{split}\\
		\partial_{z}g= & \Gamma\left(g_{0}-g\right)-g\left|E\right|^{2}\,,\label{eq:Haus2}\\
		\partial_{z}q= & q_{0}-q-sq\left|E\right|^{2}\,.\label{eq:Haus3}
\end{align}
Here, $z$ and $\xi$ denote the fast and slow time scales which describe the evolution within one round-trip and from one round-trip to the next one, respectively. Time is normalized to the absorber recovery time, $g_0$ is the pumping rate, $\Gamma$ is the gain recovery rate, $q_0$ is the value of the unsaturated losses that determines the modulation depth of the saturable absorber. Further, the gain bandwidth is $\gamma$, the cavity losses are $k$ and $\alpha_{g,q}$ correspond to the line-width enhancement factors for semiconductor material, while the ratio of the saturation energy of the gain and the absorber media is $s$. Note that the HME~(\ref{eq:Haus1})-(\ref{eq:Haus3}) was originally derived for gain media that are slowly evolving on the time scale of the cavity round-trip, resulting in a quasi-uniform gain temporal profile within the cavity. However, recent generalizations of the HME~\cite{PGG-NAC-20, HLGJ-OL-20, NV_PRE_21} preserve carrier memory from one round-trip toward the next, allowing for the correct description of complex pulse trains such as harmonic ML regimes. Since we operate in the long cavity regime without the loss of generality we apply periodic boundary conditions.
The phase and translational invariance of the electromagmetic field in the passively mode-locked laser operated in the long cavity regime allow each pulse $j\in[1,\,N]$ to possess different phases $\varphi_j$ and positions $z_j$, see Fig.~\ref{fig:sketch} for $N=5$. The lasing threshold $g_{\mathrm{th}}$ is determined by the pump level $g_0$ for which the off solution $(E,\,g,\,q)=(0,\,g_0,\, q_0)$ becomes linearly unstable and is defined by $g_\mathrm{th}=q_0+2k$. Since we operate in the regime of temporal localized states, we choose $g_0<g_{\mathrm{th}}$. We note that in our formalism weak time-delayed feedback appears as a non-local spatial term with amplitude $\eta$ and phase $\Omega$ in the field equation.

\begin{figure}
	\centering
	\includegraphics[width=1\columnwidth]{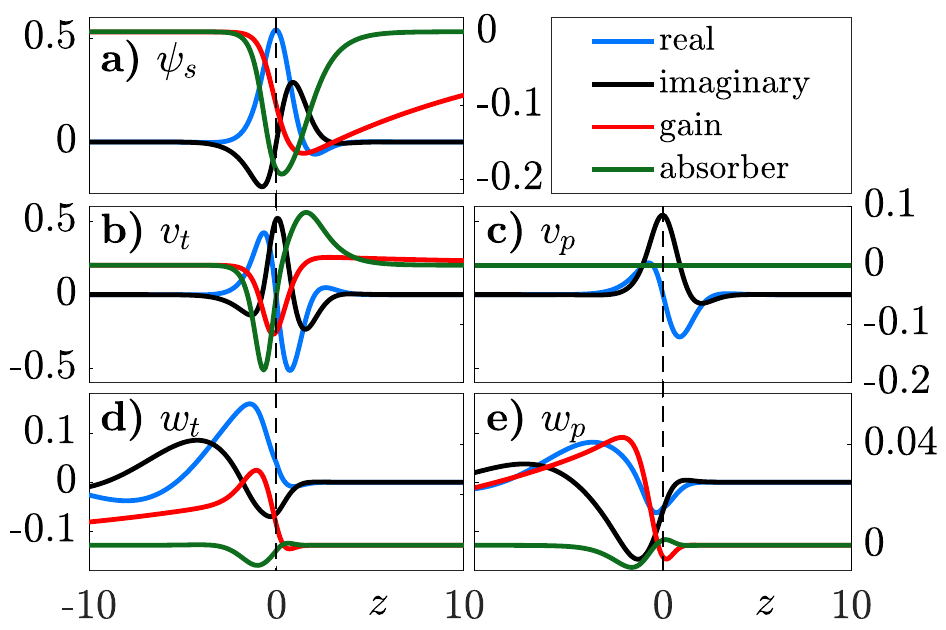}
	\caption{Exemplary stationary profiles of the (a) field and carrier components $\psi_s$, (b) translation and (c) phase neutral eigenfunctions and (d) translational (e) phase neutral eigenfunctions of the adjoint eigenvalue problem. For all panels, the left and right axis stands for the field and the carrier components, respectively. Parameters are $(\gamma,k,\alpha_g,\alpha_q,g_0,\Gamma,q_0,s)=(5,0.1056,1.5,0.5,0.99g_\text{th},0.08,0.268,10)$ in a domain of length $\tau=200$. We use a discretization of 2048 points.}
	\label{fig:profiles}
\end{figure}

One can readily solve the HME~(\ref{eq:Haus1})-(\ref{eq:Haus3}) numerically in the long cavity regime with periodic boundary conditions using, e.g., the pseudo-spectral split-step semi-implicit scheme presented in Appendix \ref{sec:numerics-Haus} while ensuring to use a fine enough mesh in order to prevent effects imposed by the discretization \cite{CER-PTRSA-11}. From that, one can extract the dynamical information for the position and the phases of the resulting pulses.
However, both the phases and the positions of the pulses evolve on a slow timescale which would require long time simulations over $10^5$ of round-trips. Hence, in order to access the effective information of $\varphi_j$ and $z_j$ and the pulse interactions, we derive the \emph{effective equations of motion}. The EOM provide a general interaction law between $N$ pulses in the cavity and is achieved by projecting the high-dimensional dynamics of the HME~(\ref{eq:Haus1})-(\ref{eq:Haus3}) onto the low-dimensional subspace of the slowly evolving phase and translation modes. We rewrite the HME~(\ref{eq:Haus1})-(\ref{eq:Haus3}) for a real-valued vector function $\psi=\left(\mathrm{Re}(E),\,\textrm{Im}(E),\,g-g_0,\,q-q_0\right)^T$ yielding
\begin{equation}
 M\partial_{\xi}\psi=\mathcal{L}\psi+\mathcal{N}(\psi)+\eta e^{i\Omega}M\psi(z-\tau_f)\,, \label{eq:HME_gen}
\end{equation}
where $M=\mathrm{diag}(1,1,0,0)$ is a mass matrix and $\mathcal{L}$ and $\mathcal{N}$ are linear and nonlinear operators, respectively. Defining
$$\psi_s=\psi_{s,f}+\psi_{s,c}=\left(\mathrm{Re}(E_s),\,\textrm{Im}(E_s),\,g_s-g_0,\,q_s-q_0\right)^T$$
as a stationary solution of Eq.~\eqref{eq:HME_gen}, which is composed of the stationary solution for the field components $\psi_{s,f}=\left(\mathrm{Re}(E_s),\,\textrm{Im}(E_s),\,0,\,0\right)$ and the carrier components $\psi_{s,c}=\left(0,\,0,\,g_s-g_0,\,q_s-q_0\right)$, we can employ the ansatz that a multi-pulse state $\Psi$ is the sum of multiple stationary pulses $\psi_s$ which are shifted and rotated with respect to each other. We denote the shifts $z_j$ and rotations $\varphi_j$, respectively (cf. blue and orange arrows in Fig.~\ref{fig:sketch}):
\begin{equation}
	\Psi=\sum_{j=1}^{N}\psi_{j}+\delta\psi\,,
	\label{eq:ansatz}
\end{equation}
where $\delta \psi$ accounts for small deviations from the simple summation and
\begin{equation}
	\psi_{j}=R\left(\varphi_j\right)\psi_s\left(z-z_j\left(\xi\right)\right)\,,
	%\left(\begin{array}{c}
	%	e^{i\varphi_{j}\left(\xi\right)} \psi_{s,f}\left(z-z_{j}\left(\xi\right)\right)\\
	%	\psi_{s,c}\left(z-z_{j}\left(\xi\right)\right)
	%\end{array}\right)\,.
\end{equation}
where $R\left(\varphi\right)$ is a matrix that rotates the field components of $\psi_s$ by a phase $\varphi$.
The four profiles of the vector function $\psi_s$ calculated for a typical parameter set are shown in Fig.~\ref{fig:profiles}~(a).
Plugging the ansatz~(\ref{eq:ansatz}) into Eq.~\eqref{eq:HME_gen} and linearizing around the stationary solution $\psi_s$ leads to the linear eigenvalue problem  $\mathcal{J}_{s}v=M\,\lambda v$ where $\mathcal{J}_{s}$ is the Jacobian. The latter possesses two neutral eigenvalues corresponding to two neutral eigenfunctions $v_t=\partial_{z}\psi_{s}$ and $v_p=\partial_{\varphi}\psi_{s}$ for the translational and phase invariance, respectively, see Fig.~\ref{fig:profiles} (b), (c). In the next step, we calculate the eigenmodes $w_t$ and $w_p$ of the corresponding adjoint linear system defined by $\mathcal{J}_{s}^{\dagger}w=M\,\bar{\lambda} w$, see  Fig.~\ref{fig:profiles} (d), (e) for the corresponding profiles.

\begin{figure}
	\centering
	\includegraphics[width=1\columnwidth]{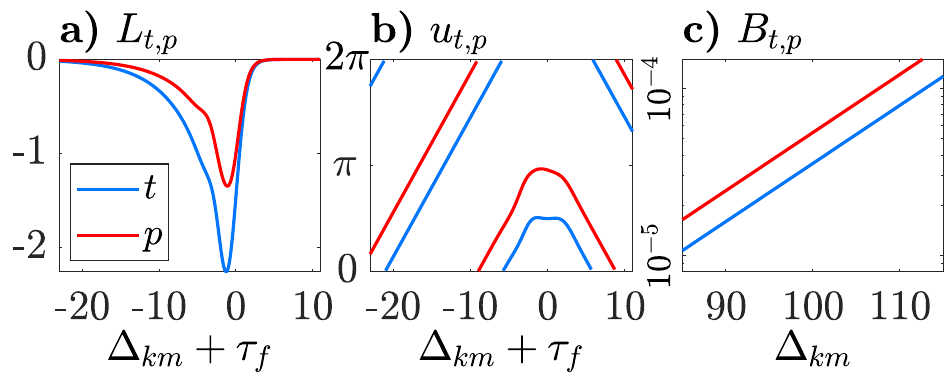}
	\caption{Numerically computed nonlinear force contributions $L_{t,p}$, $u_{t,p}$ and $B_{t,p}$ of EOM~(\ref{eq:EOM1}), (\ref{eq:EOM2}) as a function of $\Delta_{km}$.  All the forces are periodic functions with a periodicity of $\tau$. Note, that panel (c) is given on a logarithmic scale. Parameters are as in Fig.~\ref{fig:profiles}. Formal definitions given in App.~\ref{sec:App}, Eqs.\,(\ref{eq:Haus_EOM_l}),\,(\ref{eq:Haus_EOM_phase_u}) and (\ref{eq:Haus_EOM_tail1}), respectively. }
	\label{fig:fit funs}
\end{figure}

The core of the EOM derivation consists in noticing that the perturbation $\delta \psi$ in Eq.~(\ref{eq:ansatz}) must remain small and bounded. As such, all the terms in the EOM must be orthogonal to the kernel of $\mathcal{J}_{s}$ since, otherwise, $\delta \psi$ would grow without bound.
This condition is achieved by projecting onto the adjoint eigenmodes $w_t$ and $w_p$ and using that $(v_i,w_i)$ form a biorthogonal set. We further assume that the pulses are well-separated so that the interactions are weak, and that we can neglect shape deformation modes. Finally, note that we are considering the solution basis for the case of zero feedback. This approach allows us to add weak feedback as a source of motion and phase dynamics. In other words, the time-delayed feedback induced satellites are treated as a first order perturbation. In practice, it means that the feedback induced satellites are small enough that they do not induce any  interaction with the carriers. This leads to the following EOM (see Appendix~\ref{sec:App} for the complete derivation):

\begin{align}
	\begin{split}
		\partial_{\xi}\varphi_{m}=	\sum_{k\ne m}&\big[B_{p}\left(\Delta_{km}\right)+\eta L_{p}\left(\tau_{f}+\Delta_{km}\right)\times \\
	&\sin\left(\Omega+\theta_{km}+u_{p}\left(\tau_{f}+\Delta_{km}\right)\right)\big],
	\end{split} \label{eq:EOM1}\\
	\begin{split}
		\partial_{\xi} z_{m}=	\sum_{k\ne m}&\big[B_{t}\left(\Delta_{km}\right)+\eta L_{t}\left(\tau_{f}+\Delta_{km}\right)\times\\
	&\sin\left(\Omega+\theta_{km}+u_{t}\left(\tau_{f}+\Delta_{km}\right)\right)\big]\,.
	\end{split}\label{eq:EOM2}
\end{align}
Here, we define the phase and position differences $\theta_{km}\equiv \varphi_k-\varphi_m$ and $\Delta_{km}\equiv z_k-z_m$. Further, $B_{t,p}\left(\Delta_{km}\right), L_{t,p}\left(\Delta_{km}\right)$ and $u_{t,p}\left(\Delta_{km}\right)$ are nonlinear functions, giving the information about the interaction strength. These forces result from overlap integrals between the adjoint eigenvectors and the solution profile (see App.~\ref{sec:App}, Eqs.\,(\ref{eq:Haus_EOM_l}),\,(\ref{eq:Haus_EOM_phase_u}),\,(\ref{eq:Haus_EOM_tail1})  for the formal definition) and their numerically computed profiles are depicted in Fig.~\ref{fig:fit funs}. We note, that we neglect the phase-dependent force stemming from the overlapping tails (cf. term with coefficient $A_{t,p}$ in Eqs. (\ref{eq:Haus_EOM6},\ref{eq:Haus_EOM7})). Assuming exponential tails, this interaction scales as $\exp(-\Delta_{km}/\tau_p)$ where $\tau_p$ is the pulse width. With $\tau_p\approx2$ (cf. Fig.~\ref{fig:profiles}(a)) and $\Delta_{km}\approx100$, this force is extremely weak compared to the other relevant forces (cf. Fig. \ref{fig:fit funs}).
That is, the relevant forces stem from gain repulsion and interaction via time-delayed feedback, respectively.

While Eqs.~(\ref{eq:EOM1}),\,(\ref{eq:EOM2}) consider interactions between all pulses, generally nearest neighbor interactions provide the leading order term. Note that the efficiency of the EOM is that the overlap integrals only need to be evaluated once for a particular set of system parameters. They can then be applied independently of the pulse number and the feedback parameters $\eta$, $\Omega$ and $\tau_f$.

Analyzing the general structure of the EOM~(\ref{eq:EOM1}),\,(\ref{eq:EOM2}) one notices that both the position and phase terms are composed of two parts: a phase-independent term $B_{t,p}$ (cf. Fig.~\ref{fig:fit funs}~(c)) stemming from interaction between pulses via carriers (which only depend on the intensity) 
and the presence of non-zero linewidth enhancement factors, and a phase-dependent term stemming from the interaction via time-delayed feedback. In fact, the phase-independent force $B_{t,p}$ can be calculated analytically. It is defined as (cf. App.~\ref{sec:App})
\begin{align}
	B_{t,p}\left(\Delta\right)=\frac{1}{N_{t,p}}\left\langle w_{t,p}\vert\mathcal{J}_{s}\psi_{s,c}\left(z-\Delta\right)\right\rangle\,,
	\label{eq:B force}
\end{align}
where $N_{t,p}$ are the normalization factors for the translation and phase part, respectively, which read
\begin{equation*}
 N_p =\left\langle w_{p}\vert Mv_{p}\right\rangle\,,\quad N_t =-\left\langle w_{t}\vert Mv_{t}\right\rangle\,,
\end{equation*}
and $\langle\cdot|\cdot\rangle$ is a scalar product. First, as the gain recovers much slower than the absorber ($\Gamma\ll 1$), we can neglect the attracting effect of the saturable absorber such that the phase-independent force is governed by gain repulsion~\cite{V-Optics-23}. We define $z_{0}$ as the position, where the pulse has decayed sufficiently such that one can assume exponential recovery. Hence, we obtain
\begin{align}
	%G\left(x\right)	=\left(G_{s}-G_{0}\right)e^{-\Gamma\left(x-x_{0}\right)}\equiv Te^{-\Gamma x}.
	\psi_{s,g}\left(z\right)=T
	\begin{cases}
		e^{-\Gamma z} & z>z_{0}\\
		e^{-\Gamma (z+\tau)} & z<z_{0}
	\end{cases}\,,
	\label{eq:psiG}
\end{align}
 where $T=[g\left(z_0\right)-g_{0}]e^{\Gamma z_{0}}$. The factor of $e^{-\Gamma \tau}$ in the second equation has to be added due to the periodicity of the domain. From Eq.~(\ref{eq:psiG}) one can see that $\psi_{s,g}(z-\Delta)= e^{\Gamma\Delta}\psi_{s,g}(z)$ which can be used to solve the scalar product in Eq.~\ref{eq:B force}. One obtains
\begin{align}
 &B_{t,p}\left(\Delta\right)=C_{t,p} e^{\Gamma \Delta}\,, \label{eq:B force2}\\
 \text{where }\quad&C_{t,p} =\frac{T}{N_{t,p}} \left\langle {w}_{t,p}\vert\mathcal{J}_{s}\left(e^{-\Gamma\left(z+\tau\right)}e_{g}\right)\right\rangle\,.\nonumber
\end{align}
Here, $e_g$ is a unit vector with components in the gain part. For more details, see App.~\ref{subsec:Analytical-carrier_interaction}. Hence, we see that as the carrier recovers exponentially, the phase-independent force is an exponential as well. We note that the attracting interaction via the absorber also follows an exponential and can be derived in the same way~\cite{V-Optics-23}. The interaction scales as $\exp(-1\cdot\Delta_{km})$, where the coefficient of 1 stems from the normalization of Eqs.~(\ref{eq:Haus1}-\ref{eq:Haus3}) with respect to the absorber recovery rate. For separations of $\Delta_{km}\approx100$ which we consider in the following, the interaction is $\sim 10^{-44}$ which safely can be neglected compared to the gain interaction which is $\sim10^{-4}$.
While the exponential interaction in eq. (\ref{eq:B force2})  can be understood intuitively for the dynamics of the pulses' positions $z_j$, it can be surprising that the phases $\varphi_{j}$ are also influenced by a phase-independent force. The answer to this lies in the line-width enhancement factors $\alpha_{g,q}$ which couple the real and imaginary parts of the electric field, i.e. for $\alpha_{g,q}=0$ we find $B_p\left(\Delta\right)=0$.

\begin{figure*}
	\centering
	\includegraphics[width=1\textwidth]{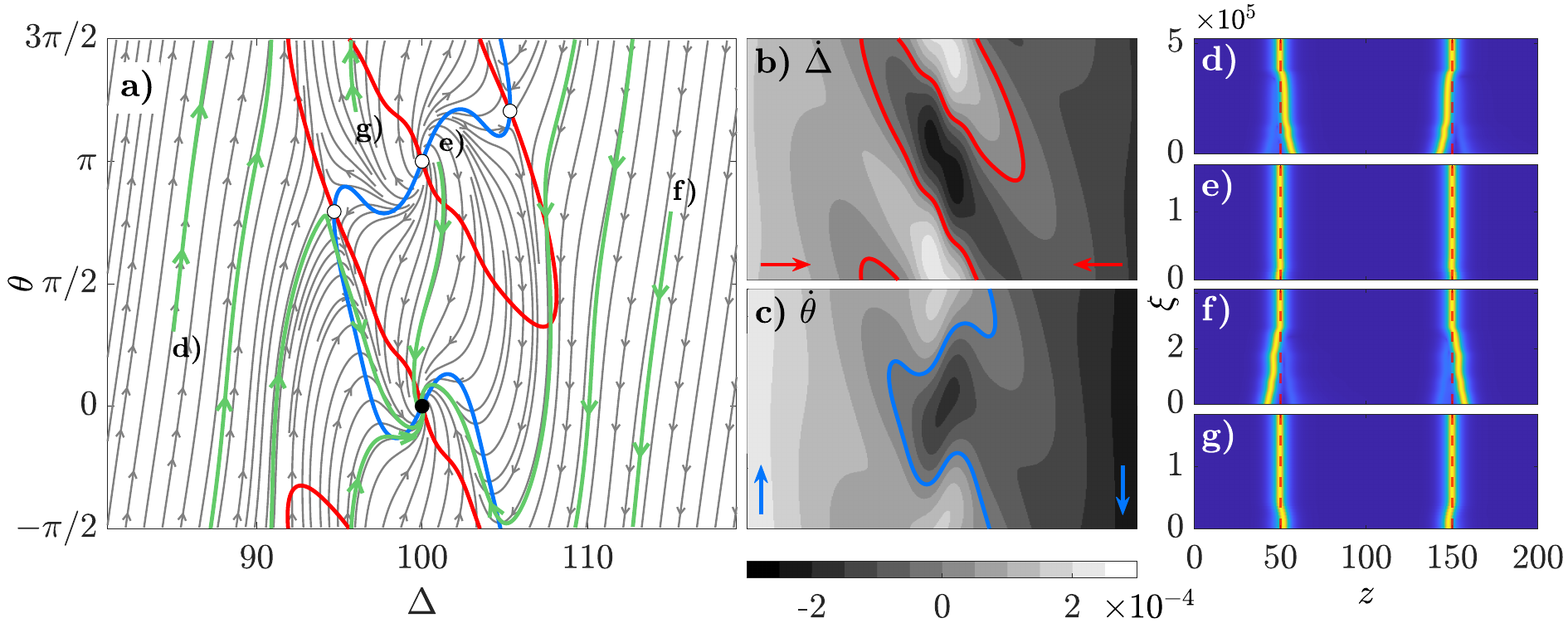}
	\caption{(a) Vector field diagram of the phase space $(\Delta,\,\theta)$ governed by the EOM (\ref{eq:EOM1}),(\ref{eq:EOM2}) for two pulses. Red and blue contour lines correspond to the $\Delta$- and $\theta$-nullclines, respectively. Black (white) circles at the intersections of the nullclines correspond to stable (unstable) fixed points. Light green lines correspond to four exemplary trajectories obtained from time simulations of the full HME~(\ref{eq:Haus1})-(\ref{eq:Haus3}). The respective two-time representations of the intensities are shown in (d)-(g). There,  vertical red dashes lines indicate the equidistant position, i.e., $z_1=50$, $z_2=150$. (b),(c) The magnitude of $\dot \Delta$ and $\dot \theta$, respectively. (d)-(g): Four exemplary time traces corresponding to the four initial states in (a). To emphasize the presence of the satellite, the colormap corresponds to $\left|E\right|^{1/4}$. Parameters are: $\tau_f=100,\eta=10^{-4}$,$\Omega=\pi/4$, other parameters as in Fig.~\ref{fig:profiles}.}
	\label{fig:flow EOM}
\end{figure*}

In order to visualize the derived EOM~(\ref{eq:EOM1}),\,(\ref{eq:EOM2}), we start with a simple configuration where two pulses with the positions $z_{1,2}$ and phases $\varphi_{1,2}$ circulate in the cavity, leading to a two-dimensional phase space spanned with one position and one phase difference $\Delta_{21}=\Delta= z_2-z_1$ and $\theta_{21}=\theta=\varphi_2-\varphi_1$, respectively. Here, the flow is described by the two ODEs $\dot \Delta\left(\Delta,\theta\right)$ and $\dot \theta\left(\Delta,\theta\right)$ which follow directly from the EOM~(\ref{eq:EOM1}),\,(\ref{eq:EOM2}) (cf. App. \ref{sec:App_two_pulses}, Eqs.\, (\ref{eq:Haus_EOM_Delta}),\,(\ref{eq:Haus_EOM_theta})). First, it is instructive to look at symmetries in the resulting equations. From Eqs.~(\ref{eq:EOM1}),\,(\ref{eq:EOM2}) we can directly see, that $\dot \Delta \left(\Delta,\theta\right) = - \dot \Delta \left(-\Delta,-\theta\right)$ (the same property holds for $\dot \theta$). This property can be concluded from the exchanging the numbering of the two pulses. Further, $\dot \Delta$ and $\dot \theta$ are $\tau$- and $2\pi$-periodic in the first and second variable, respectively. Therefore, one obtains
$$
\dot \Delta \left(\Delta,\theta\right) = - \dot \Delta \left(\tau-\Delta,2\pi-\theta\right).
$$
Having that in mind,  we can directly deduce two steady states: $(\Delta^*,\theta^*)=\left(\frac{\tau}{2},\pi\right)$ and $(\Delta^*,\theta^*)=\left(\frac{\tau}{2},0\right)$. That is, at $\Delta=\frac{\tau}{2}\pm\delta$, $\theta=\pi\pm\alpha$ one obtains
\begin{align}
	\dot \Delta \left(\frac{\tau}{2}+\delta,\pi+\alpha\right) &= - \dot \Delta \left(\frac{\tau}{2}-\delta,\pi-\alpha\right). \label{eq:symmetry}
\end{align}
Next, for the sake of simplicity, we start with the case of resonant feedback, i.e., the situation when the pulse satellite of a pulse induced by the feedback loop coincides with another pulse, i.e., $\tau_f=\tau/2$ for $\tau=200$. The other parameters are set to $\Omega=0.25 \pi$ and $\eta=10^{-4}$ such that the interactions via time-delayed feedback and carriers are in the same order of magnitude (cf. $y$-scales in Fig.~ \ref{fig:fit funs}). For each point in the phase space $(\Delta,\,\theta)$, one can evaluate the flux of the EOM~(\ref{eq:EOM1}),(\ref{eq:EOM2}) and reconstruct the corresponding vector field, see Fig. ~\ref{fig:flow EOM}~(a).
Due to the aforementioned symmetries, we find the one stable  $(\Delta^*,\,\theta^*)=(\tau/2,\,0)$ (filled circle) and one unstable $(\Delta^*,\,\theta^*)=(\tau/2,\,\pi)$ (open circle) fixed point, corresponding to the equidistant configurations. Furthermore, we find two other steady states corresponding to non-equidistant configurations. In Fig. \ref{fig:flow EOM}~(a) they are placed at around $\Delta\approx95$ and $\Delta\approx105$, however, they are both unstable.

At this point, we want to emphasize the excellent quantitative predictions by the EOM. For that, we performed four DNSs (cf. black labels (d)-(g) in  Fig.~\ref{fig:flow EOM}~(a) for the corresponding initial conditions) of the full HME~(\ref{eq:Haus1})-(\ref{eq:Haus3}) with two pulses, see also Fig.\ref{fig:flow EOM}~(d)-(g) for the two-time representation of the corresponding time traces. Here, vertical red lines visualize the equilibrium positions at $z_1=50$ and $z_2=150$ for $\tau=200$ and $\tau_f=100$. Note that in panels (d), (f) the effect of the satellites due to time-delayed feedback is especially visible: at the steady state, the one pulse is sitting on the satellite of the second one and vise versa.

From the resulting time trace, the positions and phases of the pulses are extracted and plotted as trajectories in the phase space (cf. green lines in Fig.~\ref{fig:flow EOM}~(a)). One can see that not only stable steady states are predicted correctly, but in the whole phase space the trajectories are tangent to the predicted vector field. Note that as the present interaction forces are small, the simulations of the HME (\ref{eq:Haus1})-(\ref{eq:Haus3}) need to performed over several $10^5$ round-trips. This reveals a strong advantage of the EOM~(\ref{eq:EOM1}),\,(\ref{eq:EOM2}) as the vector field for the complete phase space can be reconstructed in seconds versus hours it would take to achieve a similar result with the full HME model.

\begin{figure}
	\centering
	\includegraphics[width=1\columnwidth]{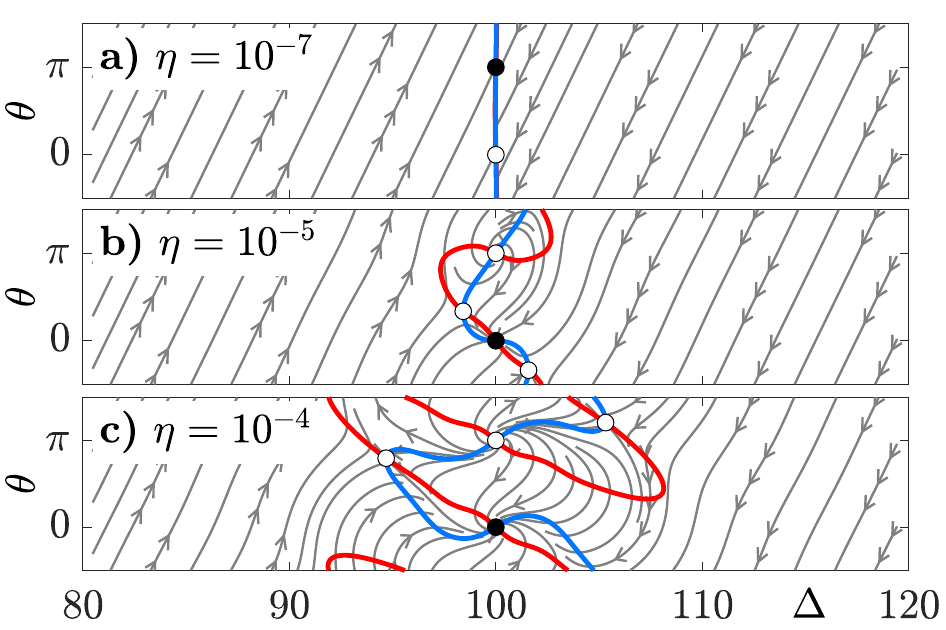}
	\caption{Vector field diagrams of the phase space $(\Delta,\,\theta)$ governed by the EOM~(\ref{eq:EOM1}),\,(\ref{eq:EOM2}) for two pulses for three different values of $\eta$. Red and blue contour lines correspond to the $\Delta$- and $\theta$-nullclines, respectively. Black (white) circles at the intersections of the nullclines correspond to stable (unstable) fixed points. Other parameters as in Fig.~\ref{fig:flow EOM}.}
	\label{fig:flow etas}
\end{figure}

In this case the feedback delay $\tau_f=100$ was chosen such that it mainly acts around the equidistant position at $\Delta=100$. The width of the area where the feedback has a strong influence is given by the width of the amplitude of the phase-dependent force  $L_{t,p}$ (cf. Fig.~\ref{fig:fit funs}~(a)). In regions, where the influence of the feedback is small ($\Delta \lesssim 90$ and $\Delta\gtrsim110$), the system is driven by gain repulsion. This force pushes the pulses to equidistance, hence, it points towards the center (cf. red arrows in Fig. \ref{fig:flow EOM}~(b)), where the magnitude of $\dot{\Delta}$ is shown together with its nullcline (red contour line). At the same time, due to the carrier frequency, the phase changes at a similar rate (cf. blue arrows and $\dot \theta$ nullcline in Fig.~\ref{fig:flow EOM}~(c)), resulting in the effective diagonal motion in the phase space (cf. Fig. \ref{fig:flow EOM}~(a)).

\begin{figure*}
	\centering
	\includegraphics[width=1\textwidth]{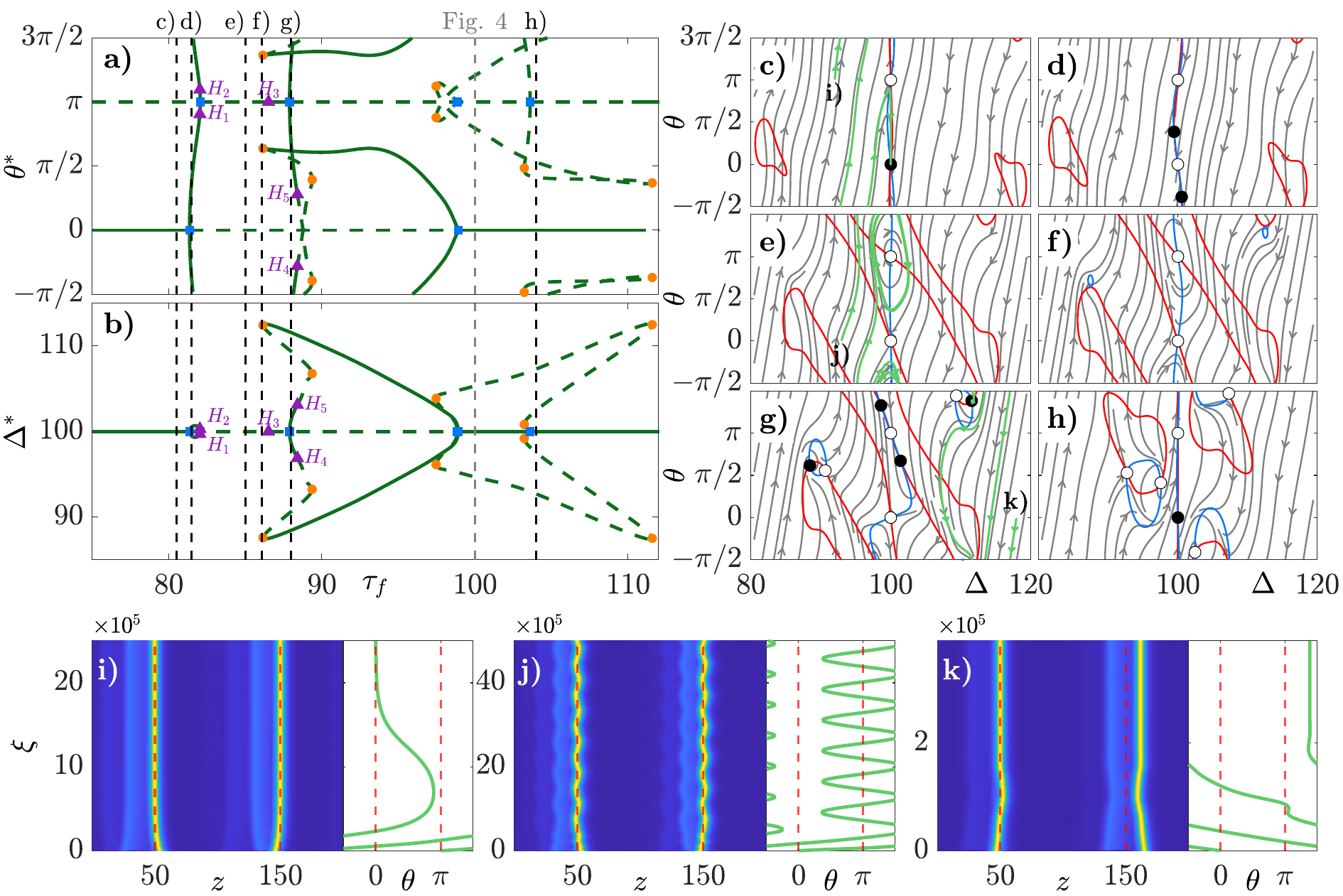}
	\caption{(a),(b) Bifurcation diagrams in $\tau_f$ of the EOM~ (\ref{eq:EOM1}),(\ref{eq:EOM2}) for $\eta=10^{-4}$ and $\Omega=\pi/4$ showing steady states $(\theta^*,\,\Delta^*)$ in phase and position differences, respectively. Solid (dashed) green lines stand for stable (unstable) unstable solutions. Purple triangles, blue squares and orange dots correspond to AH, pitchfork and saddle-node bifurcation points, respectively. (c)-(h) Vector field diagrams for values of $\tau_f$ indicated by black dashed lines in panels (a) and (b). Red and blue contour lines correspond to the $\dot\Delta$- and $\dot\theta$-nullclines, respectively. Black (white) circles at the intersections of the nullclines correspond to stable (unstable) fixed points. See the supplemental material for a scan through the bifurcation diagram. (i)-(k) Result of time simulations of the full HME~(\ref{eq:Haus1})-(\ref{eq:Haus3}) where equidistant positions and phases are indicated by red dashed lines. The corresponding trajectories are shown in panels (c), (e) and (g), respectively. Other parameters as before.}
	\label{fig:bif diag}
\end{figure*}

For the results presented in Fig.~\ref{fig:flow EOM}, the value of $\eta$ was chosen in a way that the phase-dependent force is in the same order of magnitude as the phase-independent force. A situation, when this is not the case can be seen in Fig.~\ref{fig:flow etas}, where the phase space $(\Delta,\,\theta)$ together with the corresponding nullclines is presented for three values of $\eta$. One can see in Fig.~\ref{fig:flow etas}~(a), that for smaller values of $\eta$, the phase space is solely dominated by the force due to gain repulsion, which results in diagonal vector field lines towards the equidistant positions as discussed above. For larger values of $\eta$, more complex dynamics can appear as seen in  Fig. \ref{fig:flow etas} (b), (c). Note there is an upper bound for what values of $\eta$ provide good predictions as in the derivation of the EOM~(\ref{eq:EOM1}),\,(\ref{eq:EOM2}), it is assumed, that the satellite stemming from the time-delayed feedback loop does are not sufficiently intense to create a carrier depletion. While this is true for small values of $\eta$, this is not the case for large $\eta$.

In order to gain a full overview of the phase and position dynamics in dependence of the feedback parameters $\tau_f$ and $\Omega$, we perform a detailed bifurcation analysis of the EOM~(\ref{eq:EOM1}),(\ref{eq:EOM2}). It is conducted by calculating the vector field as was presented in Fig.~\ref{fig:flow EOM}, extracting the steady states $(\Delta^*,\,\theta^*)$ from it and calculating  its behavior by changing the control parameters.

The results of a scan in feedback delay time $\tau_f$ are presented in Fig.~\ref{fig:bif diag}~(a),(b), where the equilibrium phase $\theta^*$ and position $\Delta^*$ differences are shown, respectively. Note that the stable (solid line) and unstable (dashed line) steady states at $\theta^*=(0,\,\pi)$ in the panel (a) lie on top of each other in the panel (b) as they both correspond to the equidistant case $\Delta^*=100$. Further, different phase space configurations, corresponding to different $\tau_f$ values labeled with the vertical black dashed lines in Fig.~\ref{fig:bif diag}~(a),(b) are shown in Fig.~\ref{fig:bif diag}~(c)-(h). Furthermore, the gray dashed line indicates the $\tau_f=100$-cut previously presented in Fig.~\ref{fig:flow EOM}~(a).

One can see that the bifurcation diagrams in Fig.~\ref{fig:bif diag}~(a),(b) reveal a complex dynamics with many interesting regimes. First, at low values of $\tau_f$, there are only the two fixed points corresponding to the in-phase and anti-phase equidistant configurations (cf. filled and open circles in Fig.~\ref{fig:bif diag}~(c), where both $\Delta$ (red) and $\theta$ (blue) nullclines cross). This means that the satellites are too far away from the pulses for them to interact. This is emphasized by the green trajectory corresponding to the result of a time simulation where the pulses settle on an equidistant configuration (cf. also Fig.~\ref{fig:bif diag}~(i)). However, in the final state pulses and satellites exhibit small overlap which is in contrast to the situation shown in Fig.~\ref{fig:flow EOM}~(d)-(g). This results in a de facto decoupling of phase and position dynamics as the pulses quickly settle on an equidistant state, yet the phases take much longer to reach the steady state.

Next, by increasing $\tau_f$, two non-equidistant configurations emerge from the in-phase solution $\theta^*=0$ (cf. Fig.~\ref{fig:bif diag}~(d)) in a pitchfork bifurcation (cf. blue square in Fig.~\ref{fig:bif diag}~(a)). Further, on these branches, two Andronov-Hopf (AH) bifurcations denoted $H_1$ and $H_2$ occur leading to the emergence of time-periodic solutions. In fact, in the range of $\tau_f\approx[82,\,86]$, all fixed points are unstable such that the limit cycle is the only stable manyfold. This can be seen in the phase-space in Fig.~\ref{fig:bif diag}~(e). Here, the green line is a trajectory obtained from a time simulation of the full HME~(\ref{eq:Haus1})-(\ref{eq:Haus3}) which confirms the presence of a stable periodic orbit. The two-time representation of the trace is shown in Fig.~\ref{fig:bif diag}~(j). Again, the quantitative agreement between the HME and the EOM is excellent.

\begin{figure}
	\centering
	\includegraphics[width=1\columnwidth]{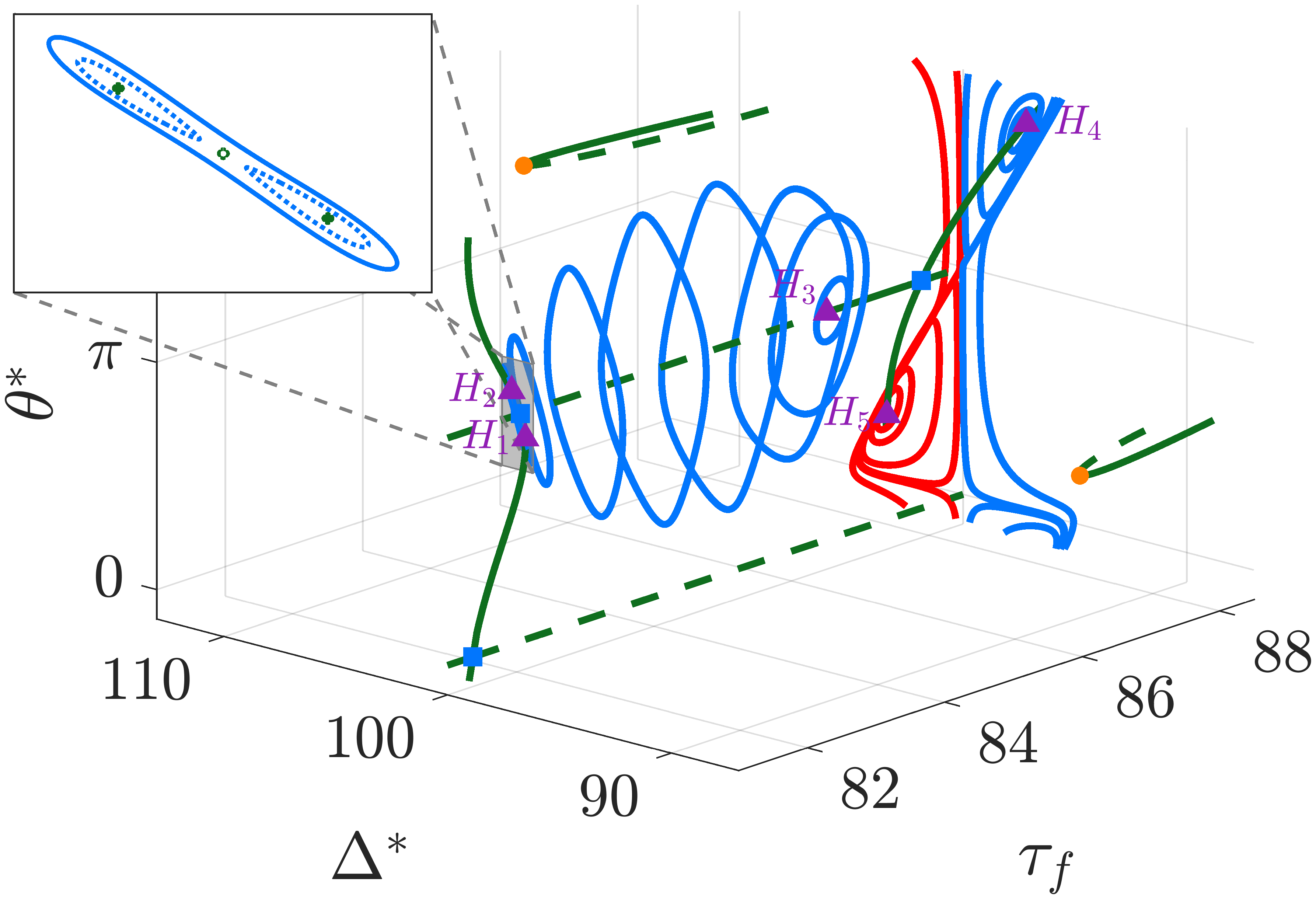}
	\caption{Three-dimensional phase space spanned by $(\tau_f,\,\Delta^*,\,\theta^*)$ of the EOM (\ref{eq:EOM1}),(\ref{eq:EOM2}). Steady states are depicted in green and emerging periodic solutions in blue and red. Purple triangles, blue squares and orange dots correspond to AH, pitchfork and saddle-node bifurcation points, respectively. The inset shows a single stable limit cycle, resulted from a double-homoclinic bifurcation followed by a saddle-node bifurcation. For a two dimensional projection, see Fig. \ref{fig:bif diag}~(a),(b).}
	\label{fig:POs}
\end{figure}

To clarify the question, why only one stable limit circle emerges in time-simulations, we consider the three-dimensional space $(\Delta^*,\,\theta^*,\,\tau_f)$ in an interval $\tau_f\approx[80,\,88]$, see Fig.~\ref{fig:POs}. A detailed bifurcation analysis reveals that the AH bifurcations $H_{1,2}$ are subcritical and the two unstable periodic orbits connect in a double-homoclinic bifurcation leading to a single unstable limit cycle. The latter undergoes a saddle-node bifurcation resulting in a single stable limit cycle (cf. inset in Fig.~\ref{fig:POs}). Note that the emerging periodic orbit oscillates around the equidistant steady state $(\Delta^*,\theta^*)=(100,\pi)$. Further, the periodic branch merge with the equidistant steady state branch again in the subcritical AH bifurcation $H_3$ around $\tau_f\approx86$. Furthermore, around $\tau_f\approx 88.5$, we find another regime of periodic orbits. Here, they bifurcate at points $H_{4,5}$ from two branches emerging from a pitchfork bifurcation of the equidistant steady state (cf. Fig.~\ref{fig:POs}). However, the AH bifurcations $H_4$ and $H_5$ are supercritical, i.e. one finds stable periodic orbit oscillating around the non-equidistant steady states. However, with increasing $\tau_f$, these limit cycles from both sides recombine at the unstable equidistant steady states in homoclinic bifurcations such that the periodic orbits now perform full phase rotations of $2\pi$.

Bringing our attention back to Fig.~\ref{fig:bif diag} after the regime of periodic solutions, we find a broad range of stable non-equidistant configurations. Figure~\ref{fig:bif diag}~(f) indicates the presence of two saddle-node bifurcations (cf. orange circles in Fig.~\ref{fig:bif diag}~(a),(b) and Fig.~\ref{fig:POs}), leading to the emergence of fixed point pairs. This corresponds to situations where one pulse sits on the satellite of the other pulse acting as an anchor, while its satellite does not interact with the other pulse, i.e. the coupling is unidirectional; cf. the time simulation in Fig.~\ref{fig:bif diag}~(k). As the anchor satellite moves linearly with $\tau_f$ we can see an approximately linear response of the steady state position $\Delta^*$ in this regime. A particularly interesting situation can be seen in Fig.~\ref{fig:bif diag}~(g), where four stable non-equidistant solutions coexist. The regime of stable non-equidistant states disappears in a pitchfork bifurcation (cf. blue square in Fig.~\ref{fig:bif diag}~(a),(b) around $\tau_f\approx100$) as the difference between the equidistant state and the time-delay $\tau_f$ approaches the pulse width. Then the satellite is absorbed into the main pulse which can be seen in form of a pitchfork bifurcation.

Finally, considering time delays larger than $\tau/2$, we do not observe other stable steady states than the equidistant ones, see Fig.~\ref{fig:bif diag}~(h). This is because the interaction of the pulses with satellites is much stronger on the leading than on the trailing edge. As a result, the pulses are almost decoupled in this regime and hence, the equidistant steady states are only weakly stable and unstable, respectively. The fact that the interaction via satellites is not symmetrical around the pulse can already be observed in the form of the amplitude of force $L_{t,p}$ in Fig.~\ref{fig:fit funs}~(a) which has a much larger tail on the leading side than on the trailing edge. This property is inherited from the asymmetric pulse profile shown in Fig.~\ref{fig:profiles}~(a).

\begin{figure}
	\centering
	\includegraphics[width=1\columnwidth]{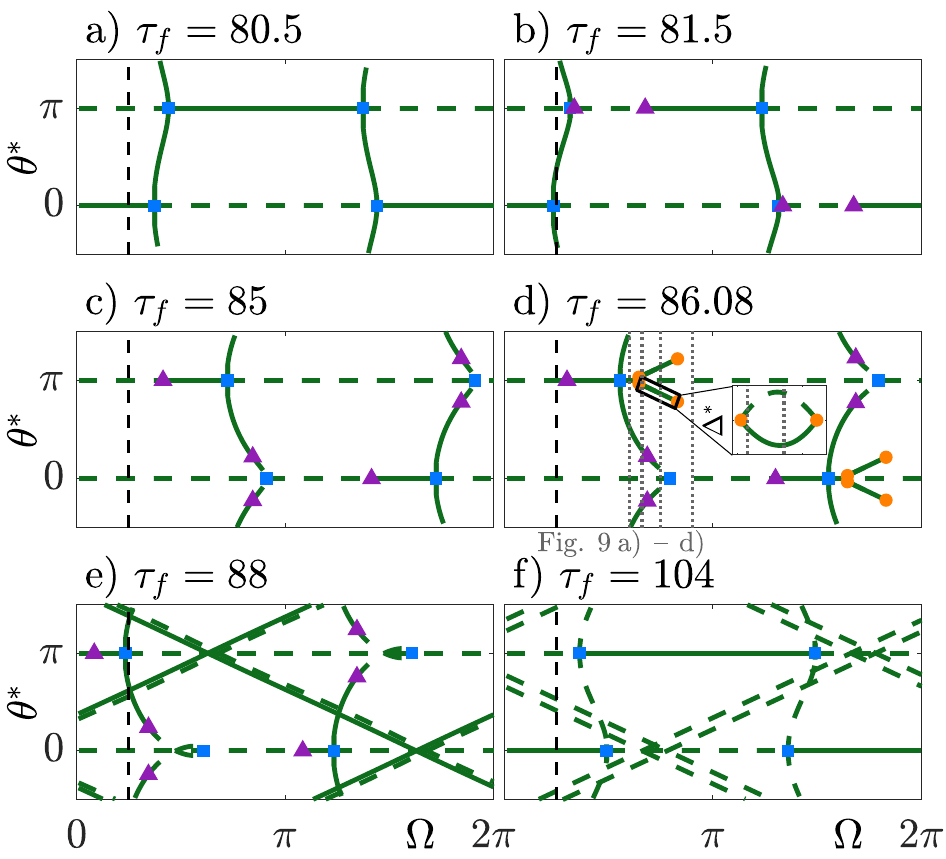}
	\caption{Bifurcation diagrams of the steady state $\theta^*$ as function of $\Omega$ for different values of $\tau_f$ (the same values as in Fig.~\ref{fig:bif diag} (c)-(h) are used). The black dashed line at $\Omega=\pi/4$ marks the cross-sections shown in Fig. \ref{fig:bif diag}~(c)-(h). In panel (d), the gray dashed lines correspond to the vector field diagrams presented in Fig.~ \ref{fig:emergence bistab}. The inset shows the appearance of the saddle-node bifurcation of steady states which lie inside the black rectangle on the position norm. See the supplemental material for scans through the shown bifurcation diagrams.}
	\label{fig:bif diag Omega}
\end{figure}

Next, we consider the influence of the feedback phase $\Omega$ on possible pulse configurations. First, we notice a further symmetry in the EOM~(\ref{eq:EOM1},\ref{eq:EOM2}):
\begin{align}
	\dot\Delta\left(\Delta,\theta,\Omega\right)=\dot\Delta\left(\Delta,\theta+\pi,\Omega+\pi\right).
	\label{eq:sym in Omega}
\end{align}
The same relation holds for $\dot \theta$. The symmetry can be observed in Fig. ~\ref{fig:bif diag Omega}, where bifurcation diagrams in $\Omega$ for various values of $\tau_{f}$ are shown. In all panels the black vertical dashed lines correspond to the cuts at $\Omega=\pi/4$ corresponding to the phase spaces shown in Fig. \ref{fig:bif diag}~(c)-(h). In Fig.~\ref{fig:bif diag Omega}~(a), one can see that when pulse and satellite are far from each other, stable non-equidistant states resulting from the pitchfork bifurcations (cf. blue squares) can only occur for a very narrow range of the feedback phases as interactions are weak in this regime. Slightly increasing $\tau_f$ (cf. panel (b)) enables the emergence of AH bifurcations (see the purple triangles) leading to periodic solutions whose range in $\Omega$ increases for larger values of $\tau_f$ (cf. panel (c)). Next, increasing $\Omega$, we find multistabilty of non-equidistant states as two pairs of saddle-node bifurcations (cf. orange circles panel (d)) emerge (cf. the inset).

\begin{figure}
	\centering
	\includegraphics[width=1\columnwidth]{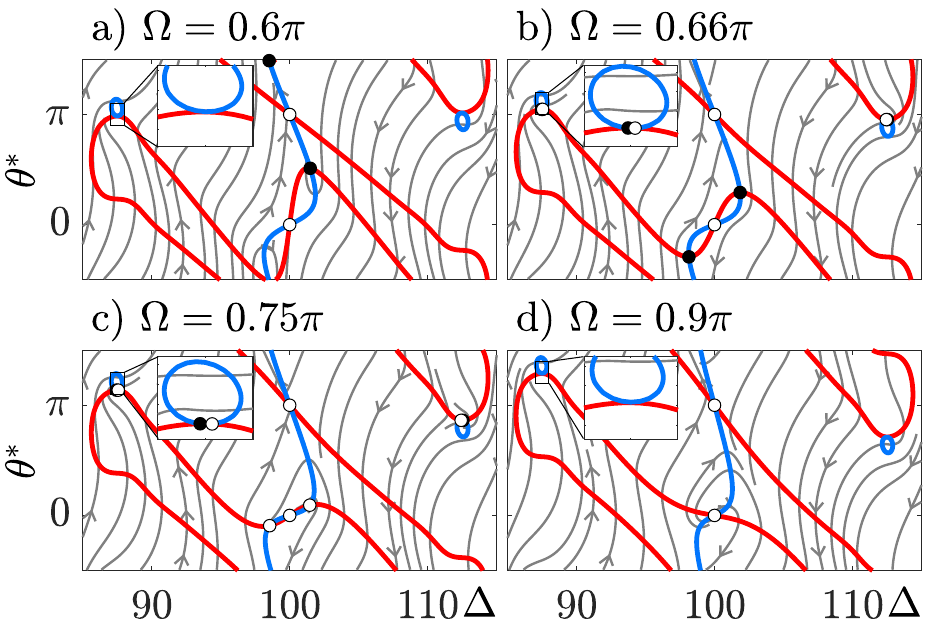}
	\caption{ Vector field diagrams evaluated at positions marked in Fig.~
	\ref{fig:bif diag Omega}~(d) by gray dashed lines. Red and blue contour lines correspond to the $\Delta$- and $\theta$-nullclines, respectively. Black (white) circles at the intersections of the nullclines correspond to stable (unstable) fixed points. The insets show close-ups on the area at around $\Delta\approx 88$, where a pair of fixed points emerges (b,c) and vanishes (a,d) in a saddle-node bifurcation. Due to the symmetry around $\Delta=100$, a pair of fixed points emerges simultaneously at around $\Delta\approx112$. }
	\label{fig:emergence bistab}
\end{figure}

To further understand this emergence, we consider the phase-space $(\Delta,\,\theta)$ for four values of the feedback phase in Fig.~ \ref{fig:emergence bistab}. There, one can observe the nullclines of phase- and position-dynamics to be very close (cf. the inset in panel (a)). Next, in panels (b) and (c), they intersect which explains the emergent stable and unstable steady states (cf. the insets). This fixed points exists far away from equidistant states  and corresponds to the situation where one pulse sits on satellite of the second pulse, while the satellite of the second pulse does not interact with the first pulse. Finally, the nullclines separate again (cf. panel (d)). Interestingly, in Fig.~\ref{fig:bif diag Omega}~(e) the aforementioned non-equidistant configuration exists for all $\Omega$ as becomes evident by the emerging cross-shaped structure of the trajectories. Further, this shape reveals that the pulse's phase relation in this case depends almost perfectly linearly on $\Omega$. At the same time, it was shown in Fig.~\ref{fig:bif diag}~(b), that in this regime the position also depends linearly on $\tau_f$, i.e. in this case the time-delayed feedback parameters directly control the pulse configuration in phase and position. Finally, in Fig.~\ref{fig:bif diag Omega}~(f) for $\tau_{f}=104$, all non-equidistant steady states became unstable and only -- depending on $\Omega$ -- the in-phase or anti-phase configurations remain stable, which, however, can be bistable for a small parameter range.

% \begin{figure}
% 	\centering
% 	%\includegraphics[width=1\columnwidth]{figs/bif_diag_Omega_v2.pdf}
% 	\includegraphics[width=1\columnwidth]{figs/bif_diag_Omega_v3.pdf}
% 	\caption{Bifurcation diagrams in $\Omega$ for different values of $\tau_f$ (the same values as in fig. \ref{fig:bif diag}c)-h) are used). The black dashed line at $\Omega=\pi/4$ marks the cross sections shown in fig. \ref{fig:bif diag}~c)-h). In panel d), the gray dashed lines correspond to the flow diagrams in fig. \ref{fig:emergence bistab}. The inset shows the steady states which lie inside the black rectangle on the position norm in order to visualize the Saddle-Node bifurcations.}
% 	\label{fig:bif diag Omega}
% \end{figure}

\begin{figure*}
	\centering
	\includegraphics[width=1\textwidth]{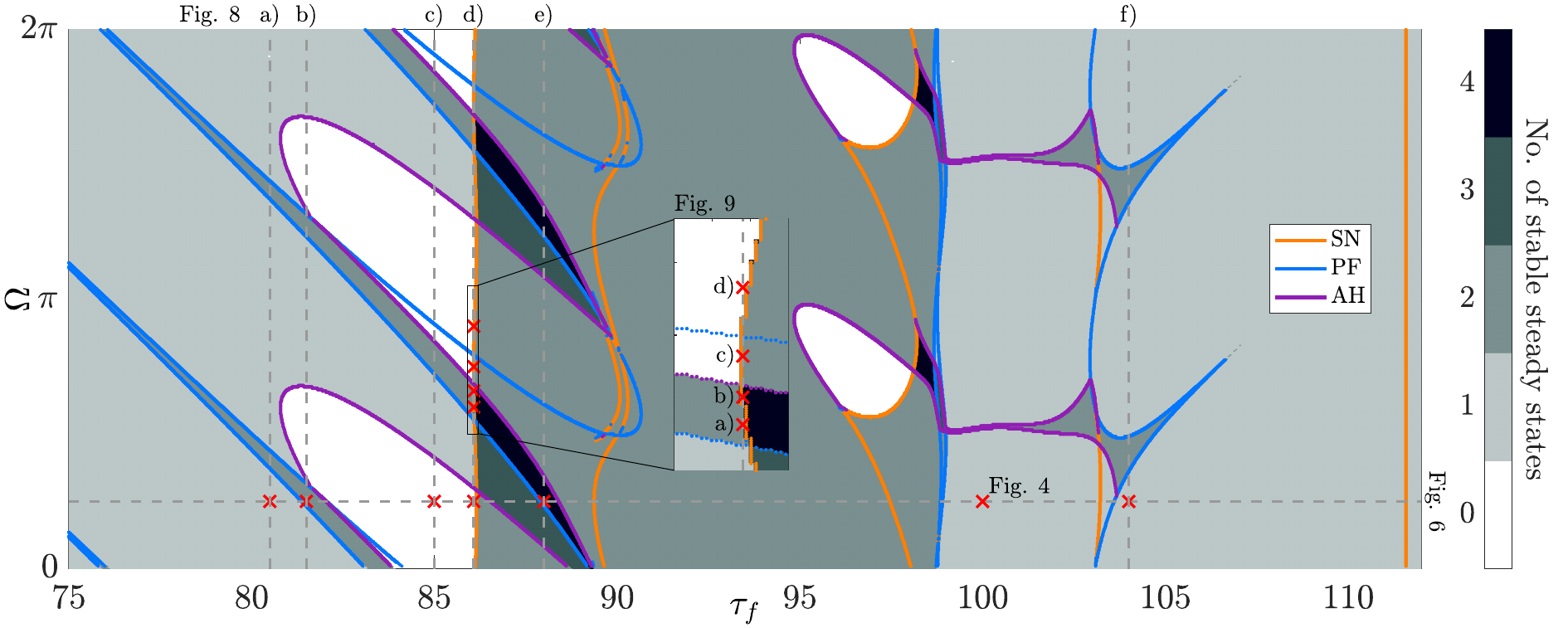}
	\caption{Two-parameter bifurcation diagram in the $(\tau_f,\Omega)$-plane. Orange, blue and purple lines stand for saddle-node (SN), pitchfork (PF)  and AH bifurcations, respectively. The colormap encodes the number of stable steady states of EOM~(\ref{eq:EOM1}),(\ref{eq:EOM2}). Dashed lines correspond to the cross-sections where one-parameter bifurcation diagrams are shown in Fig.~\ref{fig:bif diag}~(a),(b) and Fig.~\ref{fig:bif diag Omega}. At the positions marked with red crosses the phase spaces are presented in Figs. \ref{fig:flow EOM}, \ref{fig:bif diag}~(c)-(h) and Fig.~\ref{fig:emergence bistab}.}
	\label{fig:two param diag}
\end{figure*}

Finally, we summarize the findings in the two-parameter bifurcation diagram on the plane $(\tau_f,\,\Omega)$ in Fig.~\ref{fig:two param diag}. Here, the positions of all the previously shown one-parameter diagrams and corresponding phase spaces are marked by dashed lines and red crosses, respectively. Blue, orange and purple lines correspond to the pitchfork, saddle-node and AH lines. First, we notice the $\pi$-periodicity in $\Omega$ due to Eq.~(\ref{eq:sym in Omega}). Further, the colormap encodes the number of stable steady states. In white regions, where no steady state is stable, periodic solutions are the global attractors. Non-equidistant states always appear in pairs due to symmetry reasons. Hence, in regions with an odd number of stable steady states, one of the equidistant configurations must be stable. On the other hand, in regions with an even number of steady states, either both equidistant states are stable or only non-equidistant states are stable. While the aforementioned case only applies in a small range of parameter values where $\tau_f>100$ (cf. Fig. \ref{fig:bif diag Omega}~(f)), the latter regime dominates the range of $\tau_{f}\in[87,97]$.

%%%%%%%%%%%%%%%%%%%%%%%%

So far, we presented the results of a system consisting of two pulses, i.e. a two-dimensional phase space. However, the EOM~(\ref{eq:EOM1},\ref{eq:EOM2}) is valid for an arbitrary number of pulses. In fact, we can reuse the calculated profiles and eigenfunctions (cf. Fig.~\ref{fig:profiles}) and define them over a cavity of length $\tau_N = \frac{\tau}{2}N$ as they all decay to zero. Then, as the profiles and eigenmodes remain unchanged, the forces $L_{t,p}, u_{t,p}, B_{t,p}$ (cf. Fig.~\ref{fig:fit funs}) remain the same as well but just need to be defined on a larger domain. However, $L_{t,p}$ converges to zero when its argument is not in the vicinity of zero (such that it is not necessary to compute $u_{t,p}$ on a larger domain) and $B_{t,p}$ is known analytically. This enables us to predict the system's behavior even though the phase space of higher dimensional systems cannot be visualized easily.

\begin{figure}
	\centering
	\includegraphics[width=1\columnwidth]{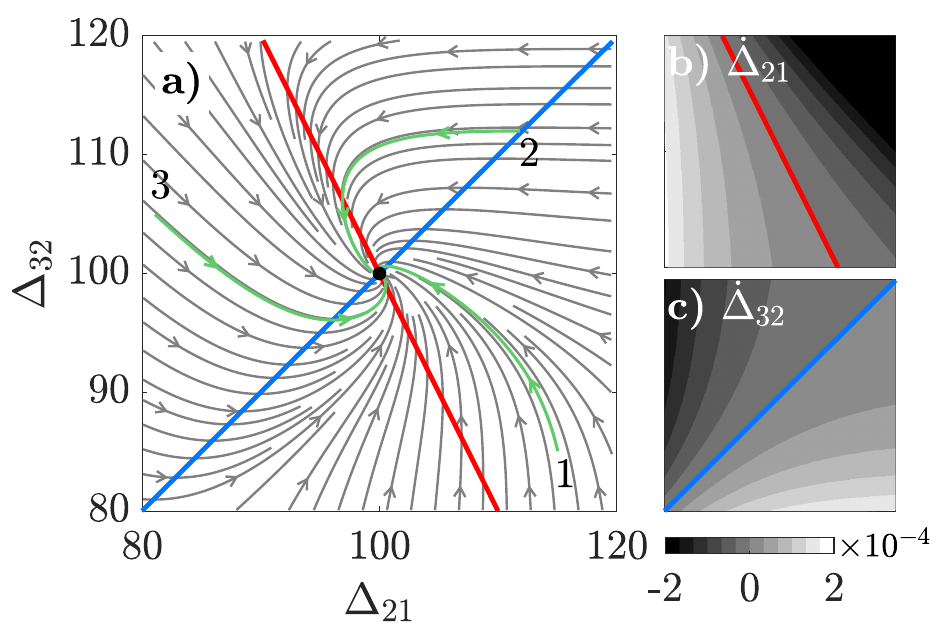}
	\caption{a) Vector field diagrams of the $(\Delta_{21},\Delta_{32})$-phase space for three pulses which dynamics is governed by the EOM  (\ref{eq:EOM1}), (\ref{eq:EOM2}) for $\eta=0$. Red and blue contour lines correspond to the $\Delta$- and $\theta$-nullclines, respectively. Black  circle at the intersections of the nullclines correspond to the stable fixed point. The green lines are traces of three numerical simulations of the full HME~(\ref{eq:Haus1})-(\ref{eq:Haus3}) and show excellent quantitative agreement with the predictions of the EOM. The magnitude of $\dot \Delta_{21}$ and $\dot \Delta_{32}$ is plotted in panels (b) and (c), respectively.}
	\label{fig:three pulses}
\end{figure}

It was demonstrated that in the range where gain repulsion is dominant the two pulses converge exponentially to the equidistant configuration (cf. Fig.~\ref{fig:flow etas}~(a)). This property can be generalized to $N$ pulses neglecting the influence of feedback ($\eta=0$). Here the gain repulsion is dominated by the first neighbor to the left, i.e. the EOM for the positions read
\begin{align}
	\dot{z}_{n}=B_t\left(z_{n-1}-z_{n}\right)
\end{align}
with the boundary conditions $z_{0}=z_{N}-\tau$ and $z_{N+1}=z_{1}+\tau$. Then the equidistant configuration is defined by $z_{n}^{*}=n\frac{\tau}{N}+v_0 t$ where $v_0$ is a constant drift. Using the ansatz $\delta_{n}=\sum_{p}a_{p}e^{\lambda_{p}\xi+iq_{p}n}$ for perturbations from these steady states (where $q_{p}=\frac{2\pi}{N}p, p=0,\dots,N-1$), one obtains (see App. \ref{sec:App_lins_stab_N_pulses} for more details)
\begin{align}
	\lambda_{p}=B_t^{\prime}\left(-\frac{\tau}{N}\right)\left[\left(\cos\left(q_{p}\right)-1\right)-i\sin\left(q_{p}\right)\right]\,.
\end{align}
As $B_t^{\prime}\left(-\frac{\tau}{N}\right)>0$ (cf. Fig.~ \ref{fig:fit funs}~(c)) and $\left(\cos\left(q_{p}\right)-1\right)<0$, the steady state is stable as expected for all $N$. However, for $N\ge3$ the imaginary part does not vanish such that these fixed points are stable spirals. To support this hypothesis we choose $N=3$ as the complete position phase space can be visualized in the $(\Delta_{21},\Delta_{32})$-plane, see Fig.~\ref{fig:three pulses}. Note that the complete phase space is four-dimensional. However, due to the absence of feedback the dynamics in the $(\theta_{21}, \theta_{32})$-plane is decoupled from the dynamics in the $(\Delta_{21},\Delta_{32})$-plane.

Here, the green lines are obtained from three time simulations of the full HME model~(\ref{eq:Haus1})-(\ref{eq:Haus3}) and show excellent quantitative agreement with the flow predicted by the EOM~(\ref{eq:EOM1}),(\ref{eq:EOM2}). Although the dynamics between pulses is given by a first order system of exponential relaxation, the ensemble give rise to oscillatory behavior and a oscillatory dynamics.

\begin{figure}
	\centering
	\includegraphics[width=1\columnwidth]{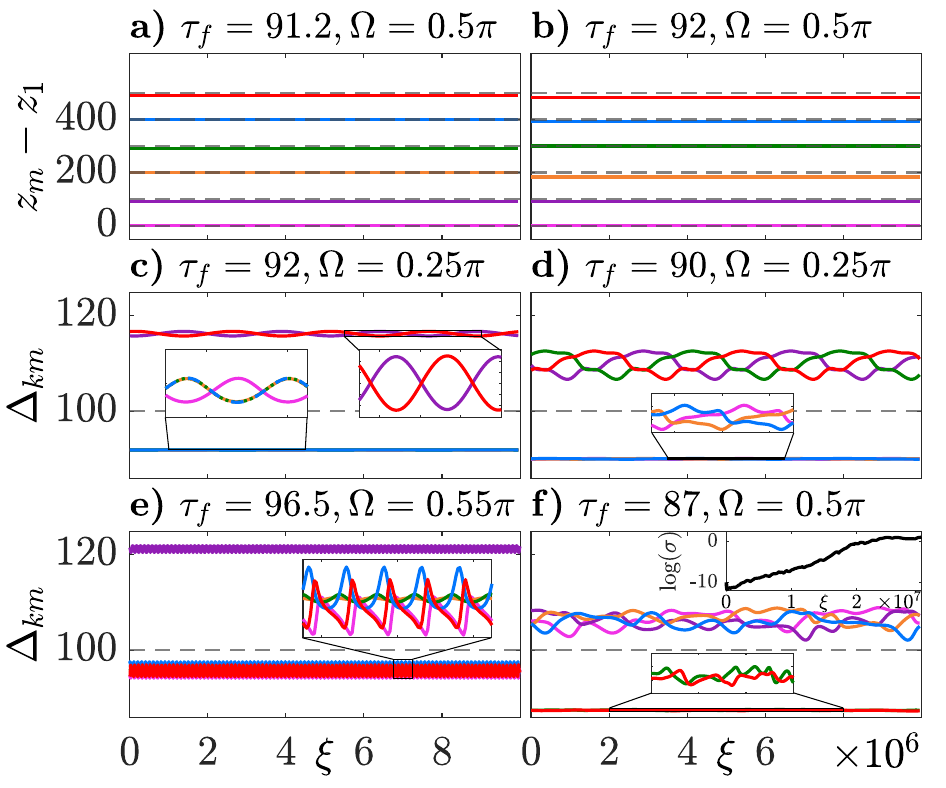}
	\caption{Time traces of numerical time simulations of EOM~(\ref{eq:EOM1}), (\ref{eq:EOM2}) for six pulses for $\eta=10^{-4}$ and for different values of $\tau_f$ and $\Omega$ yielding different configurations of phases and positions. (a) Two sets of three bound pulses. (b) Three sets of two bound pulses. (c)-(e) Different periodic states with shorter and longer periods. (f) Chaotic solution, inset: evolution of Euclidian norm $\sigma$ of perturbed and unperturbed trajectories.}
	\label{fig:HML6}
\end{figure}

Finally, including the effects of time-delayed feedback and going to even higher harmonic numbers and taking the phase dynamics into account makes the resulting dynamics more and more complex due to the added degrees of freedom. Performing various time simulations of the EOM~(\ref{eq:EOM1}),(\ref{eq:EOM2}) using different initial conditions and feedback parameters for $N=6$ reveals a variety of possible configurations as presented in Fig.~\ref{fig:HML6}. Here, panels (a) and (b) show the resulting steady states difference $z_m-z_1$ for two different $(\tau_f,\Omega)$ values, where the corresponding equidistant configurations are marked by dashed lines.

In particular, panel (a) shows a stable configuration, where the distance between neighboring pulses alternates between a higher and a lower value, i.e. there are three chunks of two bound pulses. This is analogous to the stable non-equidistant configurations found for two pulses (e.g. presented in Fig.~\ref{fig:bif diag}~(d), (g), (k)). However, in panel (b), by a slight change in $\tau_f$ we find another solution, where two chunks of three bound pulses exist.

Further, panels (c)-(e) show various periodic states. In the panel (c), we see two position differences oscillating regularly, while the remaining exhibit a much smaller amplitude (cf. the insets). Figure~\ref{fig:bif diag}~(d) presents a similar situation, however, the oscillations are not sine-like as before. Next, in panel (e), we find another periodic state which in this case has a much higher frequency, see the inset. Finally, also aperiodic oscillations can be found as shown in panel (f). Performing the simulations in panel (f) again with a small perturbation ($~10^{-10}$) and comparing the two results, reveals that the trajectories diverge exponentially. To show this, we define $\sigma$ as the Euclidian norm of the phase difference between the perturbed and unperturbed trajectories, i.e. their distance in phase space. The result is shown in the inset of Fig.~\ref{fig:HML6}~(f) where one can observe an exponential increase in $\sigma$. Hence, in this situation, the maximal Lyapunov exponent is positive and therefore, the system is chaotic.

In conclusion, we have demonstrated in this manuscript that the position and phase dynamics of a semiconductor ring laser in a multipulse configuration with $N$ pulses can be described by effective equations of motion (EOM) defined by $2N$ ODEs. In particular, we studied the interaction between pulses under the influence of time-delayed feedback which results in satellites following the main pulses. This leads to strongly non-local interactions that manifest as long-range phase-dependent forces which compete with the phase-independent gain repulsion mediated by carrier recovery and thus, enable the formation of rich dynamics.
The formulation of the problem via the EOM allowed us to perform a detailed bifurcation analysis for a two-pulse state without having to rely on the full model described by the Haus master equation. Intermediate checks of the EOM showed excellent quantitative agreement with the full model. We were able to identify many interesting regimes and, besides equidistant states, we observed non-equidistant configurations where one pulse is anchored by the satellite induced by another pulse. These non-equidistant states can also be multistable. as well as the coexistence of different non-equidistant states. 
Further, we describe periodic configurations in which the pulses oscillate around the equidistant configuration. 
Next, for states with more than two pulses, even more complex dynamics are possible due to the higher dimensionality of the problem. Besides more complex states with non-equidistant and periodic configurations described above, we also predict the existence of chaotic regimes. 
Finally, we note that in the situation where the distance between pulses gets smaller, gain repulsion becomes dominant which renders the dynamics for the relative positions very fast. This situation would allow us to perform an adiabatic elimination of these fast degrees of freedom and consider only a model for the phase of the optical pulses. This case corresponds to the model discussed in \cite{SBG-PRL24} where the interaction between the phases is described by a Kuramoto model with nearest neibhor interactions that follows from the EOM detailed here.

\section*{Declaration of competing interest}
The authors declare that they have no known competing financial
interests or personal relationships that could have appeared to influence the work reported in this paper.

\section*{Acknowledgment}
We acknowledge the financial support of the project KOGIT, Agence
Nationale de la Recherche (No. ANR-22-CE92-0009) and Deutsche Forschungsgemeinschaft
(DFG) via Grant No. 505936983. TGS and JJ acknowledge funding from
the Studienstiftung des Deutschen Volkes and the Ministerio de Economía
y Competitividad (PID2021-128910NB-100 AEI/FEDER UE, PGC2018-099637-B-100
AEI/FEDER UE), respectively.

\section*{Data availability}
Data will be made available on request.

\printcredits

%\bibliographystyle{cas-model2-names}
%\bibliographystyle{unsrtnat}

%\bibliography{full_140723,extra}



\clearpage
\appendix
\counterwithin*{equation}{section} % reset 'equation' counter whenever '\section' is executed
\renewcommand\theequation{\thesection.\arabic{equation}} % how to display the equation "number"

\section{Appendix: Derivation of the Equation of Motion}
\label{sec:App}
In this Appendix we present the derivation of the equation
of motion (EOM) starting from the Haus Master Equation (HME) Eqs. (1)-(3) of the main text.

\subsection{Definitions}

We define the scalar product in the vector space of real functions of periodicity $T$ as
\begin{equation}
	\left\langle w\vert v\right\rangle   = \int_{0}^{T}w\left(z\right)v\left(z\right)dz.\nonumber
\end{equation}
In what follows we will deal with complex functions which we define as vectors with the real and imaginary parts as components. Then the scalar product reads

\begin{equation}
	\left\langle w\vert v\right\rangle   =  \int_{0}^{T}\left[w_{x}\left(z\right)v_{x}\left(z\right)+w_{y}\left(z\right)v_{y}\left(z\right)\right]dz,\nonumber
\end{equation}
where the indices $x,y$ denote the real and imaginary parts, respectively. Now considering a linear operator $\mathcal{L}$, one can define the adjoint operator $\mathcal{L}^{\dagger}$ as
\begin{equation}
	\left\langle w\vert\mathcal{L}v\right\rangle   =  \left\langle \mathcal{L}^{\dagger}w\vert v\right\rangle .
\end{equation}
As an example, we consider the operator $\mathcal{L}=i$ (multiplication
with imaginary unit) it corresponds to a matrix and we have
\begin{align}
	\begin{split}
		i\left(\begin{array}{c}
			v_{x}\\
			v_{y}
		\end{array}\right) & =\left(\begin{array}{c}
			-v_{y}\\
			v_{x}
		\end{array}\right)\\
		&=\left(\begin{array}{cc}
			0 & -1\\
			1 & 0
		\end{array}\right)\left(\begin{array}{c}
			v_{x}\\
			v_{y}
		\end{array}\right),
	\end{split}
	\label{eq:Haus_imaginary_unit}\\
	\begin{split}
		i^{\dagger}\left(\begin{array}{c}
			v_{x}\\
			v_{y}
		\end{array}\right) & =\left(\begin{array}{c}
			v_{y}\\
			-v_{x}
		\end{array}\right)\\
		&=\left(\begin{array}{cc}
			0 & 1\\
			-1 & 0
		\end{array}\right)\left(\begin{array}{c}
			v_{x}\\
			v_{y}
		\end{array}\right),
	\end{split}\nonumber
\end{align}
such that the scalar product is preserved
\begin{align}
	\left\langle w\vert\mathcal{L}v\right\rangle  & =w_{x}\left(-v_{y}\right)+w_{y}\left(v_{x}\right)\\
	\left\langle \mathcal{L}^{\dagger}w\vert v\right\rangle  & =\left(w_{y}\right)v_{x}+\left(-w_{x}\right)v_{y}.
\end{align}

\subsection{Preliminaries}

We consider a nonlinear complex valued partial differential equation (PDE) with periodic boundaries of the form
\begin{equation}
	M\partial_{\xi}\psi=\mathcal{L}\psi+\mathcal{N}\left(\psi\right)+\eta e^{i\Omega}M\psi\left(z-\tau_{f}\right)\,.\label{eq:Haus_PDE}
\end{equation}
Here, $\mathcal{L}$ is a linear operator, $\mathcal{N}\left(\psi\right)$
is a nonlinear function and $\eta e^{i\Omega}M\psi\left(x-\tau_{f},s\right)$
a non-local term describing time-delayed feedback with the feedback delay time $\tau_f$, feedback strength $\eta$, feedback phase $\Omega$. Further, $M$ is the mass matrix and $M=\mathrm{diag}\left(1,1,0,0\right)$ for the HME in question (cf. Eqs. (1)-(3) of the main text).
Furthermore, $\psi=\psi(z,\xi)$ is a four-component real-valued vector function, and $z$ and $\xi$ denote the fast and slow time scales which describe the evolution within one round-trip and from one round-trip to the next one, respectively. The first two components of $\psi$ correspond to the real and imaginary parts of the electric
field $E$ and the third and fourth components are the gain $g$ and absorber $q$,
\begin{equation}
	\psi=\left(\begin{array}{c}
		\psi_{x}\\
		\psi_{y}\\
		\psi_{g}\\
		\psi_{q}
	\end{array}\right)=\left(\begin{array}{c}
		\mathrm{Re}\left(E\right)\\
		\mathrm{Im}\left(E\right)\\
		g-g_{0}\\
		q-q_{0}
	\end{array}\right).\nonumber
\end{equation}
Here, $g_0$ is the pumping rate and $q_0$ is the value of the unsaturated losses that determines the modulation depth of the saturable absorber.
For convenience and to account for the explicit form of the mass matrix $M$, we also define the field and carrier components as two-component vector functions
\begin{equation}
	\psi_{f}=\left(\psi_{x},\psi_{y},0,0\right)^{T},\quad\psi_{c}=\left(0,0,\psi_{g},\psi_{q}\right)^{T}\nonumber
\end{equation}
such that $\psi=\psi_{f}+\psi_{c}$.

Next, we assume that the PDE (\ref{eq:Haus_PDE})
has a stationary single-pulse solution $\psi_{s}\left(z\right)$ and two neutral modes corresponding to the phase invariance (only for the field components) and translation invariance. Hence,
every shifted and rotated version of $\psi_{s}$ is a solution of
Eq.\,(\ref{eq:Haus_PDE}) as well. Hence we define the one pulse solution $\psi_{j}=\psi_j(\xi,t)$ as
\begin{align}
	\psi_{j} & =e^{i\varphi_{j}\left(\xi\right)}\psi_{s,f}\left(z-z_{j}\left(\xi\right)\right)+\psi_{s,c}\left(z-z_{j}\left(\xi\right)\right)\nonumber\\
	& =R\left(\varphi_{j}\left(\xi\right)\right)\psi_{s}\left(z-z_{j}\left(\xi\right)\right)\,,\nonumber
\end{align}
where $\varphi_{j}\left(\xi\right)$ and $x_{j}\left(\xi\right)$
are the phase and position of the pulse which depend on the slow time,
and $\psi_{s,f}$ and $\psi_{s,c}$ are the field and carrier components
of the stationary solution. Further, $R\left(\varphi\right)$ is a
rotation matrix
\begin{equation}
	R\left(\varphi\right)=\left(\begin{array}{cccc}
		\cos\varphi & -\sin\varphi & 0 & 0\\
		\sin\varphi & \cos\varphi & 0 & 0\\
		0 & 0 & 1 & 0\\
		0 & 0 & 0 & 1
	\end{array}\right)\nonumber
\end{equation}
for the field components.
Next, we employ a superposition ansatz to describe solutions $\Psi$
that consist of $N$ pulses
\begin{equation}
	\Psi=\sum_{j=1}^{N}\psi_{j}+\delta\psi,\label{eq:Haus_superposition}
\end{equation}
where $\delta\psi$ accounts for a small deviation from the simple
summation. The aim of the following derivation is to determine how multiple
pulses $\psi_j$ interact in the cavity. This can be expressed as dynamical
\emph{equations of motion} for $\varphi_{j}\left(\xi\right)$ and $z_{j}\left(\xi\right)$,
respectively.

\subsection{Linearization}

Inserting the superposition ansatz (\ref{eq:Haus_superposition})
into Eq.\,(\ref{eq:Haus_PDE}) yields (here the dot operators on the r.h.s. denote the derivatives
with respect to slow time: $\partial_{\xi}$)
\begin{align}
	\begin{split}
		M\partial_{\xi}\Psi =\sum_{j=1}^{N}&\left[-\left(M\partial_z\psi_{j}\right)\dot{z}_{j}+\left(M\partial_\varphi\psi_{j}\right)\dot{\varphi}_{j}\right]+\\
		&M\dot{\delta\psi}
	\end{split}\nonumber\\
	\begin{split}
		=\sum_{j=1}^{N}&\mathcal{L}\psi_{j}+\mathcal{L}\delta\psi+\mathcal{N}\left(\sum_{j}^{N}\psi_{j}+\delta\psi\right)+\\
		&\eta MR\left(\Omega\right)\sum_{j=1}^{N}\psi_{j}\left(x-\tau_{f}\right)
	\end{split}\nonumber\\
	\begin{split}
		=\sum_{j=1}^{N}&\mathcal{L}\psi_{j}+\mathcal{N}\left(\sum_{j=1}^{N}\psi_{j}\right)+\\
		&\eta MR\left(\Omega\right)\sum_{j=1}^{N}\psi_{j}\left(z-\tau_{f}\right)+\\
		&\left(\mathcal{L}+\left(\frac{\partial\mathcal{N}}{\partial\psi}\right)\biggl\rvert_{\psi=\sum_{j}\psi_{j}}\right)\delta\psi,\label{eq:Haus_EOM1}
	\end{split}
\end{align}
where in the last step, we expanded the nonlinearity around $\delta\psi=0$.
Further, we can identify the Jacobian defined as
\begin{equation}
	\mathcal{J}=\mathcal{L}+\left(\frac{\partial\mathcal{N}}{\partial\psi}\right)\biggl\rvert_{\psi=\sum_{j}\psi_{j}}.\nonumber
\end{equation}
Around a single pulse, one of the $\psi_{j}$ is dominant and governs
the dynamics. Hence, the Jacobian around a pulse $\psi_{m}$ is approximately
\begin{equation}
	\mathcal{J}\approx\mathcal{J}_{m}=\mathcal{L}+\left(\frac{\partial\mathcal{N}}{\partial\psi}\right)\biggl\rvert_{\psi_{m}}.\nonumber
\end{equation}
Further, we can expand the nonlinearity in Eq.\,(\ref{eq:Haus_EOM1})
around the same pulse
\begin{equation}
	\mathcal{N}\left(\sum_{j=1}^{N}\psi_{j}\right)=\mathcal{N}\left(\psi_{m}\right)+\frac{\partial\mathcal{N}}{\partial\psi}\biggl\vert_{\psi_{m}}\sum_{j\ne m}\psi_{j}\nonumber
\end{equation}
such that one obtains
\begin{align*}
	\sum_{j=1}^{N}\mathcal{L}\psi_{j}+\mathcal{N}\left(\sum_{j=1}^{N}\psi_{j}\right) & =\left(\mathcal{L}+\frac{\partial\mathcal{N}}{\partial\psi}\biggl\vert_{\psi_{m}}\right)\sum_{j\ne m}\psi_{j}\\
	&=\mathcal{J}_{m}\sum_{j\ne m}\psi_{j}\,,
\end{align*}
where we used $\mathcal{L}\psi_{m}+\mathcal{N}\left(\psi_{m}\right)=0$
as $\psi_{m}$ solves Eq.\,(\ref{eq:Haus_PDE}) in the absence of
the feedback term. With these shortcuts, we can rewrite Eq.\,(\ref{eq:Haus_EOM1})
as
\begin{align}
	\left(M\partial_{\xi}-\mathcal{J}_{m}\right)\delta\psi=&\sum_{j=1}^{N}M\partial_z\psi_{j}\dot{z}_{j}-M\partial_\varphi\psi_{j}\dot{\varphi}_{j}+\nonumber\\
	&\mathcal{J}_{m}\sum_{j\ne m}\psi_{j}+\label{eq:Haus_EOM2}\\
	&\eta MR\left(\Omega\right)\sum_{j=1}^{N}\psi_{j}\left(z-\tau_{f}\right)\nonumber
\end{align}
which is valid around pulse $\psi_{m}$. The Jacobian $\mathcal{J}_{m}$
possess eigenvalues $\lambda_{i}^{\left(m\right)}$ corresponding to eigenfunctions
$v_{i}^{\left(m\right)}$ which are defined by the generalized eigenvalue
problem
\begin{equation}
	\mathcal{J}_{m}v_{i}^{\left(m\right)}=\lambda_{i}^{\left(m\right)}Mv_{i}^{\left(m\right)}.\nonumber
\end{equation}
We know that two (neutral) eigenvalues are zero due to
the rotational and translation symmetry in Eq.\,(\ref{eq:Haus_EOM1}).
The corresponding neutral eigenfunctions $v_{t}^{\left(m\right)}$ and $v_{p}^{\left(m\right)}$
read
\begin{align}
	v_{t}^{\left(m\right)} & =\partial_{z}\psi_{m}\label{eq:Haus_eigenfunc1}\\
	v_{p}^{\left(m\right)} & =\partial_{\varphi}\psi_{m}=iM\psi_{m}\,,\label{eq:Haus_eigenfunc2}
\end{align}
where the indices $t,p$ stand for \textit{translation} and \textit{phase},
respectively. Note that we can also define Eqs.\,(\ref{eq:Haus_eigenfunc1}),\,(\ref{eq:Haus_eigenfunc2})
in terms of the eigenfunctions of the stationary solutions $\psi_{s}$
which we simply define as $v_{t}$ and $v_{p}$ as $v_{t}^{\left(m\right)}=R\left(\varphi_{m}\right)v_{t}\left(z-z_{m}\right)$
and $v_{p}^{\left(m\right)}=R\left(\varphi_{m}\right)v_{p}\left(z-z_{m}\right)$.
In the next step, we can now decompose $\delta\psi$ in Eq.\,(\ref{eq:Haus_EOM2})
in the basis of eigenfunctions $v_{i}^{\left(m\right)}$
\begin{equation}
	\delta\psi=\sum_{j}\beta_{j}^{\left(m\right)}\left(\xi\right)v_{j}^{\left(m\right)}\,,\label{eq:Haus_expansion_eigenfuncs}
\end{equation}
where the $\beta_{j}^{\left(m\right)}\left(\xi\right)$ are the time-dependent
coefficients.

Next, we consider the corresponding adjoint eigenvalue problem defined by the adjoint operator $\mathcal{J}_{m}^{\dagger}$ which has eigenfunctions $w_{i}^{\left(m\right)}$
and eigenvalues that are the complex conjugated of the eigenvalues
of $\mathcal{J}_{m}$, i.e. 
\begin{equation}
	\mathcal{J}_{m}^{\dagger}w_{i}^{\left(m\right)}=\bar{\lambda}_{i}^{\left(m\right)}Mw_{i}^{\left(m\right)}\nonumber
\end{equation}
where we used that $M^{\dagger}=M$ for the HME. At this point it is instructive
to clarify important properties of the eigenfunctions. First, $\left(v_{i}^{\left(m\right)},w_{i}^{\left(m\right)}\right)$
form a biorthogonal set, i.e.
\begin{equation}
	\left\langle w_{i}^{\left(m\right)}\vert Mv_{j}^{\left(m\right)}\right\rangle =0,\ \forall i\ne j.\label{eq:EOM=000020biorthogonal}
\end{equation}
This property can be easily shown when computing $\left\langle w_{i}^{\left(m\right)}\vert\mathcal{J}_{m}v_{j}^{\left(m\right)}\right\rangle $
in two ways: 
\begin{align*}
	\left\langle w_{i}^{\left(m\right)}\vert\mathcal{J}_{m}v_{j}^{\left(m\right)}\right\rangle  & =\lambda_{j}^{\left(m\right)}\left\langle w_{i}^{\left(m\right)}\vert Mv_{j}^{\left(m\right)}\right\rangle ,\\
	\left\langle w_{i}^{\left(m\right)}\vert\mathcal{J}_{m}v_{j}^{\left(m\right)}\right\rangle  & =\left\langle \mathcal{J}_{m}^{\dagger}w_{i}^{\left(m\right)}\vert v_{j}^{\left(m\right)}\right\rangle\\
	&=\bar{\lambda}_{i}^{\left(m\right)}\left\langle M^{\dagger}w_{i}^{\left(m\right)}\vert v_{j}^{\left(m\right)}\right\rangle\\
	&=\bar{\lambda}_{i}^{\left(m\right)}\left\langle w_{i}^{\left(m\right)}\vert Mv_{j}^{\left(m\right)}\right\rangle .
\end{align*}
As this must hold for all $i,j$, in general $\lambda_{j}^{\left(m\right)}\ne\bar{\lambda}_{i}^{\left(m\right)},\,\forall i\ne j$
and hence $\left\langle w_{i}^{\left(m\right)}\vert Mv_{j}^{\left(m\right)}\right\rangle =0$.
Further, as the solutions $\psi_{m}$ are localized around $z_{m}$,
so are the $v_{i}^{\left(m\right)}$ and $w_{i}^{\left(m\right)}$.
As we assume well separated pulses, we obtain 
\begin{equation}
	\left\langle w_{i}^{\left(k\right)}\vert Mv_{j}^{\left(m\right)}\right\rangle \approx0,\ \forall k\ne m,\ \forall i,j.\label{eq:EOM=000020localized=000020eigenmodes}
\end{equation}

\subsection{Projection on adjoint eigenmodes}

We insert the expansion of $\delta\psi$ in the eigenfunctions defined
in Eqs.\,(\ref{eq:Haus_expansion_eigenfuncs}) into (\ref{eq:Haus_EOM2})
and identify the eigenfunctions defined in Eqs.\,(\ref{eq:Haus_eigenfunc1}),\,(\ref{eq:Haus_eigenfunc2})
to obtain 
\begin{align}
	\begin{split}
		\sum_{i}&\left(\partial_{\xi}-\lambda_{i}^{\left(m\right)}\right)\beta_{i}^{\left(m\right)}Mv_{i}^{\left(m\right)}=\\
		&\sum_{j=1}^{N}\left[Mv_{t}^{\left(m\right)}\dot{z}_{j}-Mv_{p}^{\left(m\right)}\dot{\varphi}_{j}\right]+\\
		&\mathcal{J}_{m}\sum_{j\ne m}\psi_{j}+\eta MR\left(\Omega\right)\sum_{j=1}^{N}\psi_{j}\left(z-\tau_{f}\right).\label{eq:Haus_EOM3}
	\end{split}
\end{align}
Projecting this equation onto an adjoint eigenmode $w_{k}^{\left(m\right)}$
corresponding to the same pulse and using the properties in Eqs.\,(\ref{eq:EOM=000020biorthogonal})
and (\ref{eq:EOM=000020localized=000020eigenmodes}) yields
\begin{align}
	\begin{split}
		\sum_{i}&\left(\partial_{\xi}-\lambda_{i}^{\left(m\right)}\right)\beta_{i}^{\left(m\right)}\left\langle w_{k}^{\left(m\right)}\vert Mv_{i}^{\left(m\right)}\right\rangle \\
		&=\left(\partial_{\xi}-\lambda_{i}^{\left(m\right)}\right)\beta_{k}^{\left(m\right)}\left\langle w_{k}^{\left(m\right)}\vert Mv_{k}^{\left(m\right)}\right\rangle .
	\end{split}\label{eq:Haus_dynamics_eigenmodes}
\end{align}
Inserting this into Eq.\,(\ref{eq:Haus_EOM3}) gives
\begin{align}
	\partial_{\xi}\beta_{k}^{\left(m\right)} & =\lambda_{i}^{\left(m\right)}\beta_{k}^{\left(m\right)}+C_{k}^{\left(m\right)}\,,\nonumber\\
	\text{where}\quad C_{k}^{\left(m\right)} & =\frac{\left\langle w_{k}^{\left(m\right)}\vert\text{RHS (\ref{eq:Haus_EOM3})}\right\rangle }{\left\langle w_{k}^{\left(m\right)}\vert Mv_{k}^{\left(m\right)}\right\rangle }.\nonumber
\end{align}
When projecting on the all eigenmodes except for $w_{t}^{\left(m\right)}$
and $w_{p}^{\left(m\right)}$, the $\beta_{k}^{\left(m\right)}$ will
not diverge as the corresponding eigenvalues $\lambda_{k}^{\left(m\right)}$
are negative since $\psi_{m}$ is a stable stationary solution. For
$w_{t,p}^{\left(m\right)}$ we assume the $\beta_{t,p}^{\left(m\right)}$
to evolve slowly such that $\partial_{\xi}\beta_{t,p}^{\left(m\right)}\approx0$.
Therefore we obtain
\begin{align}
	0 & =\left\langle w_{k}^{\left(m\right)}\vert\text{RHS (\ref{eq:Haus_EOM3})}\right\rangle\,.\nonumber
\end{align}
Finally, we explicitly project onto $w_{t}^{\left(m\right)}$ and
$w_{p}^{\left(m\right)}$ and use again the properties of the eigenfunctions
defined in Eqs. \,(\ref{eq:Haus_eigenfunc1}),\,(\ref{eq:Haus_eigenfunc2})
such that we obtain the EOM for the phases and positions
\begin{align}
	\begin{split}
		N_{t}^{\left(m\right)}\dot{z}_{m}  =&\sum_{j\ne m}\left\langle w_{t}^{\left(m\right)}\vert\mathcal{J}_{m}\psi_{j}\right\rangle+\\
		&\eta\sum_{j=1}^{N}\left\langle w_{t}^{\left(m\right)}\vert MR\left(\Omega\right)\psi_{j}\left(z-\tau_{f}\right)\right\rangle ,
	\end{split}
	\label{eq:Haus_EOM4}\\
	\begin{split}
		N_{p}^{\left(m\right)}\dot{\varphi}_{m}  =&\sum_{j\ne m}\left\langle w_{p}^{\left(m\right)}\vert\mathcal{J}_{m}\psi_{j}\right\rangle +\\
		&\eta\sum_{j=1}^{N}\left\langle w_{p}^{\left(m\right)}\vert MR\left(\Omega\right)\psi_{j}\left(z-\tau_{f}\right)\right\rangle\,,
	\end{split}
	\label{eq:Haus_EOM5}
\end{align}
where the normalization constants read
\begin{align}
	N_{t}^{\left(m\right)} & =-\left\langle w_{t}^{\left(m\right)}\vert Mv_{t}^{\left(m\right)}\right\rangle ,\nonumber\\
	N_{p}^{\left(m\right)} & =\left\langle w_{p}^{\left(m\right)}\vert Mv_{p}^{\left(m\right)}\right\rangle .\nonumber
\end{align}

\subsubsection{Alternative Approach using the Fredholm Alternative}

In fact, we can also use the Fredhom alternative to get to this result
faster. It states that for an inhomogeneous linear equation $Ax=b$,
one can project onto the kernel eigenmode $w$ of the corresponding adjoint homogeneous equation $A^{\dagger}w=0$ to obtain
\begin{equation}
	\left\langle w\vert b\right\rangle =\left\langle w\vert Ax\right\rangle =\left\langle A^{\dagger}w\vert x\right\rangle =0.\nonumber
\end{equation}
Hence, we consider again Eq.\,(\ref{eq:Haus_EOM2}) and use as before
that the perturbation $\delta\psi$ evolves slowly (i.e. $\partial_{\xi}\delta\psi\ll1$):
\begin{equation}
	-\mathcal{J}_{m}\delta\psi=\text{RHS (\ref{eq:Haus_EOM3})}.\nonumber
\end{equation}
As $w_{t,p}^{\left(m\right)}$ are the kernel eigenmodes of $\mathcal{J}_{m}^{\dagger}$,
we directly obtain
\begin{equation}
	0=\left\langle w_{t,p}^{\left(m\right)}\vert\text{RHS (\ref{eq:Haus_EOM3})}\right\rangle .\nonumber
\end{equation}

\subsection{The explicit form of the EOM}

In the next step, we want to evaluate the scalar products in Eqs.
(\ref{eq:Haus_EOM4}),(\ref{eq:Haus_EOM5}) to find the explicit form of the EOM.

\subsubsection{Non-local feedback term\protect\label{subsec:Haus_Non-local-term}}

First, we need to evaluate the scalar product containing the non-local term
$\psi_{j}\left(z-\tau_{f}\right)$ in Eqs.\,(\ref{eq:Haus_EOM4}),\,(\ref{eq:Haus_EOM5}).
We write the nonlocal term in terms of the stationary solution $\psi_{s}$ as
\begin{equation}
	MR\left(\Omega\right)\psi_{j}\left(z-\tau_{f}\right)=MR\left(\Omega+\varphi_{j}\right)\psi_{s}\left(z-z_{j}-\tau_{f}\right)\nonumber
\end{equation}
and use that we are dealing with period functions such that we can
shift the argument in functions of the scalar product by $z_{m}$:
\begin{align}
	&\left\langle w_{t,p}^{\left(m\right)}\vert MR\left(\Omega\right)\psi_{j}\left(z-\tau_{f}\right)\right\rangle \nonumber\\
	& =\left\langle R\left(\gamma\right)w_{t,p}\vert MR\left(\Omega+\varphi_{j}\right)\psi_{s}\left(z-z_{j}+z_{m}-\tau_{f}\right)\right\rangle \nonumber\\
	& =\left\langle R\left(\gamma\right)w_{t,p}\vert MR\left(\Omega+\varphi_{j}\right)\psi_{s}\left(z-\Delta_{jm}-\tau_{f}\right)\right\rangle\,, \label{eq:Haus_EOM_feedback1}
\end{align}
where we define $\Delta_{jm}=z_{j}-z_{m}$ and $\gamma$ is some rotation
of the $m$-th adjoint eigenfunction $w_{t,p}^{\left(m\right)}$ with
respect to the stationary state eigenfunction $w_{t,p}$. Next, we
also consider the normalization factors $N_{t,p}^{\left(m\right)}$on
the l.h.s. of Eqs.\,(\ref{eq:Haus_EOM4}),\,(\ref{eq:Haus_EOM5}) which
depend on the eigenfunctions $v_{t,p}^{\left(m\right)}$ of the $m$-th
pulse but can be shifted to depend only on the stationary modes $v_{t,p}$:
\begin{equation}
	\left\langle w_{t,p}^{\left(m\right)}\vert Mv_{t,p}^{\left(m\right)}\right\rangle =\left\langle R\left(\gamma\right)w_{t,p}\vert MR\left(\varphi_{m}\right)v_{t,p}\right\rangle .\label{eq:Haus_EOM_norm_simple}
\end{equation}
For solving Eqs.\,(\ref{eq:Haus_EOM4}),\,(\ref{eq:Haus_EOM5}) for $z_{m}$
and $\varphi_{m}$, one needs to divide the expression in Eq.\,(\ref{eq:Haus_EOM_feedback1})
with Eq.\,(\ref{eq:Haus_EOM_norm_simple}). We can multiply the ket-vectors
with $R\left(-\varphi_{m}\right)$ and bra-vectors with $R\left(-\gamma\right)$
to remove the phase dependence from the denominator:
\begin{align}
	&\frac{\left\langle R\left(\gamma\right)w_{t,p}\vert MR\left(\Omega+\varphi_{j}\right)\psi_{s}\left(z-\Delta_{jm}-\tau_{f}\right)\right\rangle }{\left\langle R\left(\gamma\right)w_{t,p}\vert MR\left(\varphi_{m}\right)v_{t,p}\right\rangle } \nonumber\\
	&=\frac{\left\langle w_{t,p}\vert MR\left(\Omega+\theta_{jm}\right)\psi_{s}\left(z-\Delta_{jm}-\tau_{f}\right)\right\rangle }{\left\langle w_{t,p}\vert Mv_{t,p}\right\rangle }\,,\label{eq:Haus_EOM_feedback2}
\end{align}
where we define the phase difference $\theta_{jm}=\varphi_{j}-\varphi_{m}$.
Next, we consider the action of the rotation operator $R\left(\alpha\right)$
which acts as
\begin{equation}
	MR\left(\alpha\right)\psi_{s}\left(z\right)=\left[\cos\left(\alpha\right)+i\sin\left(\alpha\right)\right]\psi_{s,f}\left(z\right)\,,\nonumber
\end{equation}
where the imaginary unit $i$ can be understood as the operator defined
in Eq.\,(\ref{eq:Haus_imaginary_unit}). Plugging this into the r.h.s.
of Eq.\,(\ref{eq:Haus_EOM_feedback2}) gives
\begin{align}
	\begin{split}
		&\left\langle w_{t,p}\vert MR\left(\Omega+\theta_{jm}\right)\psi_{s}\left(z-\Delta_{jm}-\tau_{f}\right)\right\rangle = \\
		& \cos\left(\Omega+\theta_{jm}\right)\left\langle w_{t,p}\vert\psi_{s,f}\left(z-\Delta_{jm}-\tau_{f}\right)\right\rangle +\\
		& \sin\left(\Omega+\theta_{jm}\right)\left\langle w_{t,p}\vert i\psi_{s,f}\left(z-\Delta_{jm}-\tau_{f}\right)\right\rangle .
	\end{split}
	\label{eq:Haus_EOM_feedback3}
\end{align}
We write out the scalar products explicitly (here the index $i$ denotes
$p$ or $t$, respectively) and assign names to them:
\begin{align}
	\begin{split}
		l_{1,i}\left(\Delta_{jm}\right) & =\left\langle w_{i}\vert\psi_{s,f}\left(z-\Delta_{jm}\right)\right\rangle\\
		&=\int \big[w_{i,x}\psi_{s,x}\left(z-\Delta_{jm}\right)+\\
		&\qquad\,\,\, w_{i,y}\psi_{s,y}\left(z-\Delta_{jm}\right)\big]dz,
	\end{split}\\
	\begin{split}
		l_{2,i}\left(\Delta_{jm}\right) & =\left\langle w_{i}\vert i\psi_{s,f}\left(z-\Delta_{jm}\right)\right\rangle\\
		&=\int\big[-w_{i,x}\psi_{s,y}\left(z-\Delta_{jm}\right)+\\
		&\qquad\,\,\, w_{i,y}\psi_{s,x}\left(z-\Delta_{jm}\right)\big]dz,\label{eq:Haus_EOM_l2}
	\end{split}\\
	\Rightarrow l_{i}\left(\Delta_{jm}\right) & =\sqrt{l_{1,i}^{2}\left(\Delta_{jm}\right)+l_{2,i}^{2}\left(\Delta_{jm}\right)}.\label{eq:Haus_EOM_l}
\end{align}
Further, we define a phase $u_{i}$ as
\begin{align}
	u_{i} & =\arctan\left(\frac{l_{1,i}}{l_{2,i}}\right),\label{eq:Haus_EOM_phase_u}\\
	\text{such that }\sin\left(u_{i}\right) & =\frac{l_{1,i}}{l_{i}},\quad\cos\left(u_{i}\right)=\frac{l_{2,i}}{l_{i}}.\nonumber
\end{align}
With all these shortcuts, we rewrite Eq.\,(\ref{eq:Haus_EOM_feedback3})
to obtain
\begin{align}
	&\left\langle w_{t,p}\vert MR\left(\Omega+\theta_{jm}\right)\psi_{s}\left(z-\Delta_{jm}-\tau_{f}\right)\right\rangle \label{eq:Haus_EOM_feedback4}\\
	&\quad=l_{t,p}\left(\tau_{f}+\Delta_{jm}\right)\sin\left[\Omega+\theta_{jm}+u_{t,p}\left(\tau_{f}+\Delta_{jm}\right)\right].\nonumber
\end{align}

\subsubsection{Tail terms}

Next, we evaluate the remaining terms in Eqs.\,(\ref{eq:Haus_EOM4}),(\ref{eq:Haus_EOM5}).
First, we express all modes in terms of the stationary modes and separate
into field and carrier components using $\psi_{s}=\psi_{s,f}+\psi_{s,c}$:

\begin{align*}
	\frac{\left\langle w_{t,p}^{\left(m\right)}\vert\mathcal{J}_{m}\psi_{j}\right\rangle }{\left\langle w_{t,p}^{\left(m\right)}\vert Mv_{t,p}^{\left(m\right)}\right\rangle }=&\frac{\left\langle w_{t,p}\vert\mathcal{J}_{s}R\left(\theta_{jm}\right)\psi_{s,f}\left(z-\Delta_{jm}\right)\right\rangle}{\left\langle w_{t,p}\vert Mv_{t,p}\right\rangle }+\\
	&\frac{\left\langle w_{t,p}\vert\mathcal{J}_{s}\psi_{s,c}\left(z-\Delta_{jm}\right)\right\rangle }{\left\langle w_{t,p}\vert Mv_{t,p}\right\rangle }\,,
\end{align*}
where $\mathcal{J}_{s}$ is the Jacobian evaluated at the stationary
solution. Here, we can already identify the phase-independent term
\begin{equation}
	b_{t,p}\left(\Delta_{jm}\right)=\left\langle w_{t,p}\vert\mathcal{J}_{s}\psi_{s,c}\left(z-\Delta_{jm}\right)\right\rangle \label{eq:Haus_EOM_tail1}
\end{equation}
from the main text. In fact, we can compute this force analytically
(cf. Sec.~\ref{subsec:Analytical-carrier_interaction}). The phase
dependent term can be handled analogously to the non-local term (cf. Sec. \ref{subsec:Haus_Non-local-term}):
\begin{align}
	\begin{split}
		&\left\langle w_{t,p}\vert\mathcal{J}_{s}R\left(\theta_{jm}\right)\psi_{s,f}\left(z-\Delta_{jm}\right)\right\rangle =\\
		& \qquad\cos\left(\theta_{jm}\right)\left\langle w_{t,p}\vert\mathcal{J}_{s}\psi_{s,f}\left(z-\Delta_{jm}\right)\right\rangle +\label{eq:Haus_EOM_tail2}\\
		& \qquad\sin\left(\theta_{jm}\right)\left\langle w_{t,p}\vert\mathcal{J}_{s}i\psi_{s,f}\left(z-\Delta_{jm}\right)\right\rangle\,.
	\end{split}
\end{align}
Again, we assign names to the integrals
\begin{align}
	a_{i,1}\left(\Delta_{jm}\right) & =\left\langle w_{i}\vert\mathcal{J}_{s}\psi_{s,f}\left(z-\Delta_{jm}\right)\right\rangle ,\nonumber\\
	a_{i,2}\left(\Delta_{jm}\right) & =\left\langle w_{i}\vert\mathcal{J}_{s}i\psi_{s,f}\left(z-\Delta_{jm}\right)\right\rangle ,\nonumber\\
	a_{i}\left(\Delta_{jm}\right) & =\sqrt{a_{1,i}^{2}\left(\Delta_{jm}\right)+a_{2,i}^{2}\left(\Delta_{jm}\right)}\nonumber
\end{align}
and the corresponding phase
\begin{align}
	r_{i} & =\arctan\left(\frac{r_{1,i}}{r_{2,i}}\right),\nonumber\\
	\text{such that }\sin\left(r_{i}\right) & =\frac{a_{1,i}}{a_{i}},\quad\cos\left(r_{i}\right)=\frac{a_{2,i}}{a_{i}},\nonumber
\end{align}
such that we can write Eq.\,(\ref{eq:Haus_EOM_tail2}) as
\begin{align}
	\begin{split}
		&\left\langle w_{t,p}\vert\mathcal{J}_{s}R\left(\theta_{jm}\right)\psi_{s,f}\left(z-\Delta_{jm}\right)\right\rangle =\\
		&\qquad\qquad a_{t,p}\left(\Delta_{jm}\right)\sin\left[\theta_{jm}+r_{t,p}\left(\Delta_{jm}\right)\right].
	\end{split}\label{eq:Haus_EOM_tail3}
\end{align}

\subsubsection{Equations of motion}

Finally, we can combine the results from the previous sections (importantly
Eqs.\,(\ref{eq:Haus_EOM_feedback4}), (\ref{eq:Haus_EOM_tail1}) and
(\ref{eq:Haus_EOM_tail3})) to obtain the EOM:
\begin{align}
	\begin{split}
		\dot{z}_{m}  =\sum_{j\ne m}&\big[ A_{t}\left(\Delta_{jm}\right)\sin\left[\theta_{jm}+r_{t}\left(\Delta_{jm}\right)\right]+B_{t}\left(\Delta_{jm}\right)+\\
		&L_{t}\left(\tau_{f}+\Delta_{jm}\right)\sin\left[\Omega+\theta_{jm}+u_{t}\left(\tau_{f}+\Delta_{jm}\right)\right]\big],
	\end{split}
	\label{eq:Haus_EOM6}\\
	\begin{split}
		\dot{\varphi}_{m}  =\sum_{j\ne m}&\big[ A_{p}\left(\Delta_{jm}\right)\sin\left[\theta_{jm}+r_{p}\left(\Delta_{jm}\right)\right]+B_{p}\left(\Delta_{jm}\right)+\\
		&L_{p}\left(\tau_{f}+\Delta_{jm}\right)\sin\left[\Omega+\theta_{jm}+u_{p}\left(\tau_{f}+\Delta_{jm}\right)\right]\big]\,,
	\end{split}
	\label{eq:Haus_EOM7}
\end{align}
where the capital letters stand for the normalized forces
\begin{equation}
	A_{t,p}=\frac{a_{t,p}}{N_{t,p}},\quad L_{t,p}=\frac{l_{t,p}}{N_{t,p}},\quad B_{t,p}=\frac{b_{t,p}}{N_{t,p}}\nonumber
\end{equation}
with the normalization factors
\begin{equation}
	N_{t}=-\left\langle w_{t}\vert Mv_{t}\right\rangle ,\quad N_{p}=\left\langle w_{p}\vert Mv_{p}\right\rangle .\label{eq:Haus_EOM_normalizations}
\end{equation}
All terms in Eqs.\,(\ref{eq:Haus_EOM6}),(\ref{eq:Haus_EOM7}) can
be understood intuitively: $A_{t,p}$ are the amplitudes of the phase-dependent force (tail interaction), $B_{t,p}$ is the amplitude of the phase-independent force (interaction via carriers) and $L_{p,t}$ are the amplitude of the force stemming from the (phase-dependent) time-delayed
feedback.

\subsection{Special cases and simplifications}

Next, we consider certain special cases and simplifications
of Eqs.\,(\ref{eq:Haus_EOM6}),\,(\ref{eq:Haus_EOM7}).

\subsubsection{Nearest neighbor interactions}

First, due to exponentially decaying tails it is sufficient to consider
only nearest neighbor interactions, i.e. the sum $\sum_{j\ne m}$
reduces to $j=m\pm1$. However, while the interaction via time-delayed
feedback is only with a single other pulse (for a single time delay),
the interaction must not necessarily be with the nearest neighbor
but depends on the relation of the time-delay $\tau$ and the length
of the feedback loop $\tau_{f}$, i.e. $j=\lfloor\frac{\tau_{f}}{\tau}N\rfloor$
(or $j=\lceil\frac{\tau_{f}}{\tau}N\rceil$, depending on what is
closer). This enables interesting scenarios, as one can create a
quasi 2D system by a proper choice of $\tau_{f}$, where pulses are
coupled in one dimension via overlapping tails and in the other dimension
via time-delayed feedback. However, here we remain with nearest neighbor
interaction and choose $\tau_{f}$ in the vicinity of $\frac{\tau}{N}$
such that $j=m-1$,
\begin{align}
	\dot{z}_{m}= & A_{t}\left(z_{m+1}-z_{m}\right)\times\nonumber\\
	&\sin\left[\varphi_{m+1}-\varphi_{m}+r_{t}\left(z_{m+1}-z_{m}\right)\right]+\nonumber\\
	&B_{t}\left(z_{m+1}-z_{m}\right)+\nonumber\\
	& A_{t}\left(z_{m-1}-z_{m}\right)\times\nonumber\\
	&\sin\left[\varphi_{m-1}-\varphi_{m}+r_{t}\left(z_{m+1}-z_{m}\right)\right]+\label{eq:Haus_EOM8}\\
	&B_{t}\left(z_{m-1}-z_{m}\right)+\nonumber\\
	& L_{t}\left(\tau_{f}+z_{m-1}-z_{m}\right)\times\nonumber\\
	&\sin\left[\Omega+\varphi_{m-1}-\varphi_{m}+u_{t}\left(\tau_{f}+z_{m-1}-z_{m}\right)\right],\nonumber\\
	\dot{\varphi}_{m}= & A_{p}\left(z_{m+1}-z_{m}\right)\times\nonumber\\
	&\sin\left[\varphi_{m+1}-\varphi_{m}+r_{p}\left(z_{m+1}-z_{m}\right)\right]+\nonumber\\
	&B_{p}\left(z_{m+1}-z_{m}\right)+\nonumber\\
	& A_{p}\left(z_{m-1}-z_{m}\right)\times\nonumber\\
	&\sin\left[\varphi_{m-1}-\varphi_{m}+r_{p}\left(z_{m+1}-z_{m}\right)\right]+\label{eq:Haus_EOM9}\\
	&B_{p}\left(z_{m-1}-z_{m}\right)+\nonumber\\
	& L_{p}\left(\tau_{f}+z_{m-1}-z_{m}\right)\times\nonumber\\
	&\sin\left[\Omega+\varphi_{m-1}-\varphi_{m}+u_{p}\left(\tau_{f}+z_{m-1}-z_{m}\right)\right]\,.\nonumber
\end{align}
In this context, we need to define the boundary conditions (i.e., what
does $\varphi_{N+1}$, $z_{N+1}$, $\varphi_{0}$ and $z_{0}$ mean).
For the positions we need to be careful and add/subtract $\tau$, i.e.,
\begin{align}
	z_{N+1} & =z_{1}+\tau,\quad z_{0}=z_{N}-\tau,\nonumber\\
	\varphi_{N+1} & =\varphi_{1},\quad\varphi_{0}=\varphi_{N}\,.\nonumber
\end{align}

\subsubsection{Damped translational mode}

Depending on the cavity geometry, the dynamics of the phases and positions
can occur on different time-scales. In particular, in short cavities,
interaction via carriers, namely repulsion via gain, is dominant,
such that the translational mode is damped and the pulses are in an
equidistant configuration $z_{m}-z_{m-1}=\frac{\tau}{N}\,\forall m$.
Hence, the coefficients of the phase-dependent terms and the phase
$r_{p}$ in Eqs.\,(\ref{eq:Haus_EOM6}),\,(\ref{eq:Haus_EOM7}) become
constants:
\begin{align}
	A_{+} & =A_{p}\left(-\frac{\tau}{N}\right),\,A_{-}=A_{p}\left(\frac{\tau}{N}\right)\nonumber\\
	r_{+} & =r_{p}\left(-\frac{\tau}{N}\right),\,r_{-}=r_{p}\left(\frac{\tau}{N}\right)\nonumber\,.
\end{align}
Inserting these relations in Eqs.\,(\ref{eq:Haus_EOM8}),\,(\ref{eq:Haus_EOM9})
and neglecting the effects of time-delayed feedback for a moment, we obtain a set of Kuramoto
equations for the phase interaction:
\begin{align}
	\begin{split}
		\dot{\varphi}_{m}=&A_{+}\sin\left(\varphi_{m-1}-\varphi_{m}+r_{+}\right)+\\
		&A_{-}\sin\left(\varphi_{m+1}-\varphi_{m}+r_{-}\right)+\omega_{0}\,,
	\end{split}
\end{align}
where 
\begin{equation}
	\omega_{0}=B_{p}\left(-\frac{\tau}{N}\right)+B_{p}\left(\frac{\tau}{N}\right)\nonumber
\end{equation}
is a common phase rotation that can be scaled out by using a comoving
reference frame.

On the other hand, when the effects of time-delayed feedback are dominant, we obtain
\begin{align}
	\begin{split}
		\dot{\varphi}_{m}=&\eta L_{p}\left(\tau_{f}-\frac{\tau}{N}\right)\times\\
		&\sin\left[\Omega+\varphi_{m-1}-\varphi_{m}+u_{p}\left(\tau_{f}-\frac{\tau}{N}\right)\right].
	\end{split}
\end{align}
An interesting special case occurs when choosing resonant feedback,
i.e. $\tau_{f}=\frac{\tau}{N}$. First, we notice, that the normalization
factor $N_{p}$ defined in Eq.\,(\ref{eq:Haus_EOM_normalizations})
is equal to the overlap integral $l_{2,p}\left(0\right)$ (cf. Eq.~
(\ref{eq:Haus_EOM_l2})). Consequently, we find 
\begin{equation}
	L_{p}\left(0\right)=\frac{l_{p}\left(0\right)}{l_{2,p}\left(0\right)}=\frac{1}{\cos u}\,,\nonumber
\end{equation}
where we used Eq.~\,(\ref{eq:Haus_EOM_phase_u}) and defined $u=u_{p}\left(0\right)$.
This leads to an effective EOM with a single free parameter $u$:
\begin{equation}
	\dot{\varphi}_{m}=\frac{\eta}{\cos u}\sin\left(\Omega+\varphi_{m-1}-\varphi_{m}+u\right).
\end{equation}

\subsubsection{Interaction of two pulses}
\label{sec:App_two_pulses}

The situation with only two pulses is special because right and left
interaction go to the same pulse. Here we define $\theta=\varphi_{2}-\varphi_{1}$
and $\Delta=z_{2}-z_{1}$ as the corresponding phase and position difference, respectively,
\begin{align}
	\begin{split}
		\dot{\Delta}= & A_{t}\left(-\Delta\right)\sin\left[-\theta+r_{t}\left(-\Delta\right)\right]+B_{t}\left(-\Delta\right)+\\
		&L_{t}\left(\tau_{f}-\Delta\right)\sin\left[\Omega-\theta+u_{t}\left(\tau_{f}-\Delta\right)\right]-\\
		& A_{t}\left(\Delta\right)\sin\left[\theta+r_{t}\left(\Delta\right)\right]-B_{t}\left(\Delta\right)-\\
		&L_{t}\left(\tau_{f}+\Delta\right)\sin\left[\Omega+\theta+u_{t}\left(\tau_{f}+\Delta\right)\right],
	\end{split}
	\label{eq:Haus_EOM_Delta}\\
	\dot{\theta}= & A_{p}\left(-\Delta\right)\sin\left[-\theta+r_{p}\left(-\Delta\right)\right]+B_{p}\left(-\Delta\right)+\nonumber\\
	&L_{p}\left(\tau_{f}-\Delta\right)\sin\left[\Omega-\theta+u_{p}\left(\tau_{f}-\Delta\right)\right]-\nonumber\\
	& A_{p}\left(\Delta\right)\sin\left[\theta+r_{p}\left(\Delta\right)\right]-B_{p}\left(\Delta\right)-\label{eq:Haus_EOM_theta}\\
	&L_{p}\left(\tau_{f}+\Delta\right)\sin\left[\Omega+\theta+u_{p}\left(\tau_{f}+\Delta\right)\right]\,.\nonumber
\end{align}
Here, we note that the forces are periodic functions, i.e. $A_{p,t}\left(-\Delta\right)=A_{p,t}\left(\tau-\Delta\right).$
From Eqs.\,(\ref{eq:Haus_EOM_Delta}), (\ref{eq:Haus_EOM_theta}),
it is easy to see that the equations are antisymmetric in $\theta$
and $\Delta$, i.e. we have
\begin{align}
	\dot{\Delta}\left(\Delta,\theta\right) & =-\dot{\Delta}\left(-\Delta,-\theta\right)\,,\nonumber\\
	\dot{\theta}\left(\Delta,\theta\right) & =-\dot{\theta}\left(-\Delta,-\theta\right)\,.\nonumber
\end{align}
This symmetry corresponds to the permutation of the pulses. Now,
we can immediately identify two steady state $\left(\Delta^{*},\theta^{*}\right)=\left(\frac{\tau}{2},0\right)$
and $\left(\Delta^{*},\theta^{*}\right)=\left(\frac{\tau}{2},\pi\right)$
corresponding to a symmetric configuration. However, the more interesting
symmetry is the one around the steady states at $\frac{\tau}{2}$.
Therefore, we consider $\Delta=\frac{\tau}{2}\pm\delta$ and $\theta=\pi\pm\alpha$ to
obtain
\begin{align}
	\dot{\Delta}\left(\frac{\tau}{2}-\delta,\pi-\alpha\right) & =-\dot{\Delta}\left(\frac{\tau}{2}+\delta,\pi+\alpha\right),\nonumber\\
	\dot{\theta}\left(\frac{\tau}{2}-\delta,\pi-\alpha\right) & =-\dot{\theta}\left(\frac{\tau}{2}+\delta,\pi+\alpha\right).\nonumber
\end{align}

\subsubsection{Linear stability analysis of $N$ equidistant pulses}
\label{sec:App_lins_stab_N_pulses}

In the absence of time-delayed feedback, we can consider a set of
$N$ pulses which interact solely via gain repulsion in a unidirectional
way (left to right), i.e.
\begin{equation}
	\dot{z}_{n}=B_{t}\left(z_{n-1}-z_{n}\right)\nonumber
\end{equation}
with the boundary conditions $z_{0}=z_{N}-\tau$ and $z_{N+1}=z_{1}+\tau$.
The stable steady state is the equidistant configuration $z_{n}^{*}=n\frac{\tau}{N}+v_{0}\xi$,
where $v_{0}=B_{t}\left(-\tau/N\right)$. To analyze its linear stability,
we consider $z_{n}=z_{n}^{*}+\delta_{n}$. Then one obtains
\begin{align}
	\dot{\delta}_{n} & =B_{t}^{\prime}\left(-\frac{\tau}{N}\right)\left(\delta_{n-1}-\delta_{n}\right)\label{eq:EOM=000020linear=000020stability=000020position}\nonumber
\end{align}
with the periodic boundary condition $\delta_{0}=\delta_{N}$ and
$\delta_{N+1}=\delta_{1}$. We expand the perturbations in Fourier
modes
\begin{equation}
	\delta_{n}=\sum_{p}a_{p}e^{\lambda_{p}\xi+iq_{p}n}\,,\nonumber
\end{equation}
where due to boundary conditions, the wave number satisfies
\begin{align}
	e^{iq_{p}N} & \stackrel{!}{=}1\nonumber\\
	\Leftrightarrow q_{p} & =\frac{2\pi}{N}p,\quad p=0,\dots,N-1.\nonumber
\end{align}
Using this expansion in Eq.\,(\ref{eq:EOM=000020linear=000020stability=000020position})
yields
\begin{equation}
	\sum_{p}a_{p}\lambda_{p}e^{\lambda_{p}\xi+iq_{p}n}=B_{t}^{\prime}\left(-\frac{\tau}{N}\right)\sum_{p}a_{p}e^{\lambda_{p}\xi+iq_{p}n}\left[e^{-iq_{p}}-1\right].\nonumber
\end{equation}
This has to be fulfilled for each element of the sum individually
because the modes are linearly independent. Therefore, we find
\begin{align*}
	\lambda_{p} & =B_{t}^{\prime}\left(-\frac{\tau}{N}\right)\left(e^{-iq_{p}}-1\right)\nonumber\\
	& =B_{t}^{\prime}\left(-\frac{\tau}{N}\right)\left[\left(\cos\left(q_{p}\right)-1\right)-i\sin\left(q_{p}\right)\right].
\end{align*}
We know that $B_{t}\left(\Delta\right)$ is exponentially recovering
with $\Delta$ (cf. Sec. \ref{subsec:Analytical-carrier_interaction})
so its derivative is positive: $B_{t}^{\prime}\left(-\tau/N\right)>0$.
Hence, we see that the steady states is indeed stable as $\cos\left(q_{p}\right)-1\le0\ \forall p$.
However, the imaginary part does not vanish for $N>2$ and therefore
leads to a convergence to the steady states in spirals. Calculating
the eigenvalues explicitly yields:
\begin{itemize}
	\item For $N=2$ we have $q_{p}=\pi p$. The corresponding eigenvalues are:
	\begin{align}
		\lambda_{0} & =0,\nonumber\\
		\lambda_{1} & =-2B_{t}^{\prime}\left(-\frac{\tau}{N}\right).\nonumber
	\end{align}
	\item For $N=3$ we have $q_{p}=\frac{2\pi}{3}p$. The corresponding eigenvalues read:
	\begin{align}
		\lambda_{0} & =0,\nonumber\\
		\lambda_{1} & =B_{t}^{\prime}\left(-\frac{\tau}{N}\right)\left[-\frac{3}{2}+i\frac{\sqrt{3}}{2}\right],\nonumber\\
		\lambda_{2} & =B_{t}^{\prime}\left(-\frac{\tau}{N}\right)\left[-\frac{3}{2}-i\frac{\sqrt{3}}{2}\right].\nonumber
	\end{align}
\end{itemize}

\subsection{Analytical approaches to carrier interaction\protect\label{subsec:Analytical-carrier_interaction}}

We want to characterize the interaction forces further. In particular, the phase-independent force reads
\begin{equation}
	B_{t,p}\left(\Delta\right)=\frac{1}{N_{t,p}}\left\langle w_{t,p}\vert\mathcal{J}_{s}\psi_{s,c}\left(z-\Delta\right)\right\rangle .\nonumber
\end{equation}
As $\Gamma\ll1$ (here the $1$ stand representative for the absorber
recovery time as this quantity is normalized), we can neglect the
effect of the saturable absorber and only consider the impact of the gain. We define $z_{0}$ as
the position where the pulse has decayed sufficiently such that one can assume exponential recovery. For $z>z_{0}$ the gain recovers as
\begin{align}
	\psi_{s,g}\left(z\right) & =\left(g_{s}-g_{0}\right)e^{-\Gamma\left(z-z_{0}\right)}=Te^{-\Gamma z},\nonumber\\
	\text{where}\quad T & =\left(g\left(z_{0}\right)-g_{0}\right)e^{\Gamma z_{0}}.\nonumber
\end{align}
We remember that this function is $\tau$ periodic. Hence, for $z<z_{0}$
the gain component reads
\begin{equation}
	\psi_{s,g}\left(z\right)=\left(Te^{-\Gamma\tau}\right)e^{-\Gamma z}.\nonumber
\end{equation}
That is, in total one obtains
\begin{equation}
	\psi_{s,g}\left(z\right)=T\begin{cases}
		e^{-\Gamma z} & z>z_{0}\,,\\
		e^{-\Gamma\left(z+\tau\right)} & z<z_{0}\,.
	\end{cases}
	\nonumber
\end{equation}
With that in mind one can find $\psi_{s,g}\left(z-\Delta\right)$. It reads
\begin{equation}
	\psi_{s,g}\left(z-\Delta\right)=Te^{\Gamma\Delta}\begin{cases}
		e^{-\Gamma z} & z>z_{0}+\Delta\,,\\
		e^{-\Gamma\left(z+\tau\right)} & z<z_{0}+\Delta\,.
	\end{cases}
	\nonumber
\end{equation}
However, we remember that for the overlap integrals we integrate over
$\left[z_{p}-a,z_{p}+a\right]$, where $a\ll\tau$ (because the Jacobian
is a local linear approximation) and $z_{p}<z_{0}$ is the position
of the pulse. As long as $\Delta>a$ (as should be the case for well-separated
pulses), we can use
\begin{equation}
	\psi_{s,g}\left(z-\Delta\right)=Te^{\Gamma\Delta}e^{-\Gamma\left(z+\tau\right)}.
	\nonumber
\end{equation}
Hence, we can calculate the force
\begin{align}
	B_{t,p}\left(\Delta\right) & =C_{t,p}e^{\Gamma\Delta}\,,\\
	\text{where }\quad C_{t,p}&=\frac{T}{N_{t,p}}\left\langle w_{t,p}\vert\mathcal{J}_{s}\left(e^{-\Gamma\left(z+\tau\right)}e_{g}\right)\right\rangle .\nonumber
\end{align}
Here, $e_{g}$ is a unit vecor with a component in the gain part.

\section{Appendix: Numerical Scheme for Integrating the Haus Master Equation}
\label{sec:numerics-Haus}
For the time simulation of the Haus master equation~(\ref{eq:Haus1})-(\ref{eq:Haus3}), we use a split-step semi-implicit method. That is, we compute nonlinearities in the original space while the linear operators are computed using spectral methods in the frequency space. Further, we need to integrate the carriers equations separately from the field because the carriers in Eqs. (\ref{eq:Haus2}),(\ref{eq:Haus3}) do not contain a temporal derivative.

%We begin to update the gain. We employ a semi-implicit step in space, which leads to the ansatz:
%\begin{align}
%	\frac{g_{i+1}^{j+1}-g_{i+1}^{j}}{\Delta z}&=\Gamma_{g}\left(g_{0}-\frac{g_{i+1}^{j+1}+g_{i+1}^{j}}{2}\right)-\frac{g_{i+1}^{j+1}+g_{i+1}^{j}}{2}I_{i}^{j}\\
%	\Leftrightarrow g_{i+1}^{j+1}&=\frac{\Delta z\Gamma_{g}g_{0}+\left[1-\frac{\Delta z}{2}\left(\Gamma_{g}+I_{i}^{j}\right)\right]g_{i+1}^{j}}{1+\frac{\Delta z}{2}\left(\Gamma_{g}+I_{i}^{j}\right)}\label{eq:carrier_update}
%\end{align}
%That is, we can compute $g_{i+1}^{j+1}$ from $g_{i+1}^{j}$. However, this implies that the scheme does not provide us with the first point $g_{i+1}^1$. For that, 
We consider the equation for the gain in Eq. (\ref{eq:Haus2}) that is the following ODE
\begin{align}
	\partial_{z}g\left(z\right)=\Gamma_{g}g_{0}-P^\prime\left(z\right)g\left(z\right)\label{eq:haus_ODE}
\end{align}
where we define 
\begin{align}
	P\left(z\right)=\int_{0}^{z}\left(\Gamma_{g}+\left|E\left(z^\prime\right)\right|^{2}\right)dz^\prime.
\end{align}
Here, we first compute the cumulative intensity
\begin{align}
	I_c\left(z\right) = \int_0^z I\left(z^\prime\right)dz^\prime
\end{align}
such that $P\left(z\right)=\Gamma_{g}z+I_{c}\left(z\right)$. The integral for $I_c\left(z\right)$ can be obtained using, e.g., the trapezoid method. Equation (\ref{eq:haus_ODE}) can be solved using the method of integrating factors. The general solution reads
\begin{align}
	g\left(z\right)=e^{-P\left(z\right)}\left(g\left(0\right)+\Gamma_{g}g_{0}\int_{0}^{z}e^{P\left(z^{\prime}\right)}dz^{\prime}\right).\label{eq:carrier_analytical}
\end{align}
Now, we apply the periodic boundary conditions, i.e., we set $g(0)=g(\tau)$ and solve for $g\left(0\right)$:
\begin{align}
	g\left(0\right)=\Gamma_{g}g_{0}\frac{\int_{0}^{\tau}e^{P\left(z^{\prime}\right)}dz^{\prime}}{e^{P\left(\tau\right)}-1}.
\end{align}
With that, we can update the carrier using the analytical expression in Eq. (\ref{eq:carrier_analytical}). The steps to obtain the updated absorber $q$ are similar.

Now for the field $E$, we denote with $E_i^j$ the $i$-th point in slow time $\xi$ and the $j$-th point in fast time $z$. We need to integrate
\begin{align}
	\partial_{\xi}E&=\left(\mathcal{L}+M\right)E,\\
	\text{where }\,\mathcal{L}&=v\partial_z+\frac{1}{2\gamma^{2}}\partial_{z}^{2}+\eta e^{i\Omega}\mathcal{S}_{-\tau_f},\\
	M&=-k+\frac{1-i\alpha_{g}}{2}g-\frac{1-i\alpha_{q}}{2}q + i\omega\,,
\end{align}
where $\mathcal{S}_{\tau}=e^{\tau\,\partial_z}$ is the shift operator, $v$ is the drift velocity and $\omega$ is the carrier frequency.
We first make an explicit half-step in time 
\begin{align}
	E_{i+\frac{1}{2}}^{j}=\left(1+\frac{\Delta t}{2}M\right)E_{i}^{j}.
\end{align}
Then, we apply the linear operator (which is done in Fourier space to handle the spatial derivatives)
\begin{align}
	\tilde{E}_{i+\frac{1}{2}}^{j}&=e^{\Delta t\mathcal{L}}E_{i+\frac{1}{2}}^{j}=\mathcal{F}^{-1}\left[e^{\Delta t\hat{\mathcal{L}}}\mathcal{F}\left[E_{i+\frac{1}{2}}^{j}\right]\right].
\end{align}
Further, we make an implicit half-step to obtain $E_{i+1}^j$:
\begin{align}
	E_{i+1}^{j}&=\frac{\tilde{E}_{i+\frac{1}{2}}^{j}}{1-\frac{\Delta t}{2}M}.
\end{align}
Finally, we consider $v$ and $\omega$. While these quantities can be scaled out of the equation, we need to find them in order to arrive at a steady state. While at the beginning of a simulation, these are additional free parameters, they can be found in a time simulation using a control loop.
For the drift $v$, we expand
\begin{align}
	I_{i}\left(z\right)=&I_{i+1}\left(z+v\Delta t\right)\\
	&=I_{i+1}\left(z\right)+\frac{dI_{i+1}}{dz}v\Delta t\,,
\end{align}
where $\frac{dI_{i+1}}{dz}=2\mathrm{Re}\left(E^*\frac{dE}{dz}\right)$. Now we weight both sides of the equation additionally with $\frac{dI_{i+1}}{dz}$ to take the tilt of the intensity profile into account. We integrate over $z$ and solve for $v$:
\begin{align}
	v=\frac{1}{\Delta t}\frac{\int_{0}^{\tau}\left(I_{i}\left(z\right)-I_{i+1}\left(z\right)\right)\frac{dI_{i+1}}{dz}dz}{\int_{0}^{\tau}\left(\frac{dI_{i+1}}{dz}\right)^{2}dz}.
\end{align}
For the carrier frequency $\omega$, we first find the phase difference as
\begin{align}
	\phi_{i+1}-\phi_{i}=\arg\left(E_{i+1}E_{i}^{*}\right).
\end{align}
But we do not simply want to sum this up because we want to the pulse itself to have a higher influence. Therefore, we have
\begin{align}
	\omega&=\frac{1}{\Delta t}\frac{\int_{0}^{\tau}\arg\left(E_{i+1}E_{i}^{*}\right)I_{i+1}dz}{\int_{0}^{\tau}I_{i+1}dz}.
\end{align}
Both equations for $v$ and $\omega$ can be understood as corrections to an already existing drift and carrier frequency, that is, they are rather $\Delta v$ and $\Delta\omega$ because $E_i$ was already computed with some $v_i$ and $\omega_i$ such that the updated quantities are $v_{i+1}=v_i+\Delta v$ and $\omega_{i+1}=\omega_i + \Delta\omega$, respectively.

\end{document}